\documentclass[prb,showpacs,preprint,endfloats]{revtex4}
\newcommand {\DM}{{\bm {D}}_{ij}}
\newcommand {\D}{{\bm {D}}}
\newcommand {\Spini}{{\bm {S}}_i}
\newcommand {\Spinj}{{\bm {S}}_j}

\newcommand {\kxy}{$\kappa _{xy}$}
\newcommand {\Tc}{$T_{\text{C}}$}
\newcommand{\LVO}{Lu$_{2}$V$_{2}$O$_{7}$}

\newcommand{\HVO}{Ho$_{2}$V$_{2}$O$_{7}$}
\newcommand{\IMO}{In$_{2}$Mn$_{2}$O$_{7}$}
\newcommand{\BMO}{BiMnO$_{3}$}
\newcommand{\YTO}{YTiO$_{3}$}
\newcommand{\LNMO}{La$_{2}$NiMnO$_{6}$}

\newcommand{\GFO}{GdFeO$_{3}$}

\usepackage{graphicx}
\usepackage{bm}
\usepackage{pifont}
\usepackage{amsmath, amssymb,amsfonts}

\begin{document}

\title{Effect of lattice geometry on magnon Hall effect in ferromagnetic insulators}

\author{T. Ideue$^1$, Y. Onose$^{1,2}$, H. Katsura$^3$, Y. Shiomi$^1$, S. Ishiwata$^1$, N. Nagaosa$^{1,4}$ and Y. Tokura$^{1,2,4}$} 
\affiliation{$^1$ Department of Applied Physics, University of Tokyo, Tokyo 113-8656, Japan \\ 
$^2$  Multiferroics Project, ERATO, Japan Science and Technology Agency (JST), Tokyo 113-8656, Japan \\
$^3$  Department of Physics, Gakushuin University, Tokyo 171-8588, Japan \\
$^4$ Cross-Correlated Materials Research Group (CMRG) and Correlated Electron Research Group (CERG), RIKEN Advanced Science Institute, Wako 351-0198, Japan
}

\date{2011}

\begin{abstract}
We have investigated the thermal Hall effect of magnons for various ferromagnetic insulators.
For pyrochlore ferromagnetic insulators Lu$_2$V$_2$O$_7$, Ho$_2$V$_2$O$_7$, and In$_2$Mn$_2$O$_7$,  finite thermal Hall conductivities have been observed below the Curie temperature \Tc . From the temperature and magnetic field dependences, it is concluded that magnons are responsible for the thermal Hall effect. The Hall effect of magnons 
can be well explained by the theory based on the Berry curvature in momentum space induced by the Dzyaloshinskii-Moriya (DM) interaction. The analysis has been extended to the transition metal (TM) oxides with perovskite structure. The thermal Hall signal was absent or far smaller in \LNMO\ and \YTO , which have the distorted perovskite structure with four TM ions in the unit cell. 
On the other hand, a finite thermal Hall response is discernible below \Tc\ in another ferromagentic perovskite oxide \BMO, which shows orbital ordering with a larger unit cell. The presence or absence of the thermal Hall effect in insulating pyrochlore and perovskite systems reflect the geometric and topological aspect of DM-induced magnon Hall effect.

\end{abstract}
\maketitle

\section{Introduction}

The Hall effect is the induction of transverse electric current upon the application of the longitudinal electric field.
While the Hall effect in nonmagnetic metals or semiconductors is usually induced by Lorentz force proportional to the magnetic field, in ferromagnets there is an additional component induced by the spontaneous magnetization, which is termed anomalous Hall effect.\cite{AHE-review} 
Recent theoretical and experimental studies show that 
the theory based on the Berry phase in momentum space accounts well for 
the observed anomalous Hall effects.\cite{AHE-review}
The spin-orbit interaction gives rise to the topological structure of the Bloch wave around band crossing points denoted as magnetic (anti-)monopoles, which act as sources or sinks of fictitious magnetic field in momentum space. The anomalous velocity caused by the fictitious field is the origin of the anomalous Hall effect. 
Because the Berry phase is not restricted to electrons, the Berry phase induced Hall effect is expected for other particles even without the charge. In fact, Hall effects of photons and phonons have been reported previously.\cite{HEL-theory,HEL-experiment,phonon-Hall-exp,phonon-Hall-theory-intrinsic,phonon-Hall-theory-extrinsic,phonon-Hall-theory3,phonon-Hall-theory4} In this paper, we study the Hall effect of magnons, which are the quanta of magnetic excitation in magnetic materials.
 
Because magnons can carry the spin moments less dissipatively than electrons do, the magnon spin current seems to be useful for future spintronics. 
In this context, the new functionalities of magnon spin currents have been investigated recently.\cite{magnon-spin-current,spin-Seebeck-insulator} The Hall effect may also be useful for the  control of the magnon spin current.
Theories of magnon Hall effect were proposed recently.\cite{magnon-Hall-theory-1,magnon-Hall-theory-2} 
Fujimoto theoretically suggested that the transverse magnon spin current can be induced by the application of the longitudinal magnetic field gradient in noncoplanar spin structure.\cite{magnon-Hall-theory-1} On the other hand, Katsura {\it et al.} showed that the ring exchange interaction induces the Hall effect even in the collinear ferromagnet in the case of particular lattice such as the Kagom$\acute{\rm e}$ lattice.\cite{magnon-Hall-theory-2} They also show that the magnon Hall effect can be observed with use of heat transport measurment, and derive a formula for the thermal Hall conductivity due to magnons.\cite{magnon-Hall-theory-2}

Quite recently, we have succeeded in 
the experimental observation of the thermal Hall response below the Curie temperature $T_{\rm C}$ in a ferromagnetic insulator with pyrochlore structure \LVO .\cite{magnon-Hall-exp}
We have found that the temperature and magnetic field dependences are consistent with the picture of magnon Hall effect.     
The observed thermal Hall conductivity can be explained by the theoretical model based on the Berry phase due to the Dzyaloshinskii-Moriya (DM) interaction.
In this paper, we have investigated the thermal Hall conductivity in various ferromagnetic insulators to further develop the concept of magnon Hall effect. We have found that the thermal Hall conductivities are commonly observed below $T_{\rm C}$ in ferromagnetic insulators with pyrochlore structure, In$_2$Mn$_2$O$_7$, Ho$_2$V$_2$O$_7$ as well as Lu$_2$V$_2$O$_7$. The temperature and magnetic field dependences for In$_2$Mn$_2$O$_7$ and Ho$_2$V$_2$O$_7$ are similar to the previously observed data of Lu$_2$V$_2$O$_7$, except for the variation of sign, indicating the generality of our observations. On the other hand, we could not observe a finite thermal Hall conductivity for the perovskite ferromagnets YTiO$_3$ and La$_2$NiMnO$_6$, in which the unit cell contains 4 magnetic transition metal (TM) sites, while a finite signal is observed for BiMnO$_3$ with a larger unit cell including 16 Mn sites. We show that these observations reflect the geometric and topological aspect of magnon Hall effect caused by the Berry phase due to the DM interaction.

The organization of the rest of the paper is as follows. 
In Sec. II, we show the details of the sample preparation and the transport measurement. 
In Sections III and IV, we discuss the thermal Hall conductivity caused by magnon Hall effect in pyrochlore and perovskite ferromagnets, respectively. Finally, we conclude with a summary in Section V.

\section{Experiment}

Crystals of \LVO\ and \YTO\ were prepared by the floating zone method.
The atmospheres were Ar gas for Lu$_2$V$_2$O$_7$ and the mixture of Ar and H$_2$ gases with the ratio of 96:4 for YTiO$_3$. The growth rate was 2 mm/h for both the cases. We prepared polycrystalline \HVO, \IMO, and \BMO\ samples by high-pressure synthesis. The mixed powder of starting material (Ho$_{2}$O$_{3}$ and V$_{2}$O$_{4}$ for Ho$_2$V$_2$O$_7$, In$_{2}$O$_{3}$ and MnO$_{2}$ for In$_2$Mn$_2$O$_7$, and Bi$_{2}$O$_{3}$ and Mn$_{2}$O$_{3}$ for BiMnO$_3$) with the prescribed ratios was packed into platinum or gold capsules ($\sim$4 mm$\phi \times$ 6 mm), and heated in a cubic anvil-type apparatus for 60 min. Samples were synthesized at 1300 $^{\circ}$C and 6.5 GPa for \HVO\ , at 850 $^{\circ}$C and 3 GPa for In$_2$Mn$_2$O$_7$, and at 700 $^{\circ}$C and 6.5 GPa for \BMO. The capsules were cooled down to room temperature before releasing the pressure. Polycrystalline \LNMO\ samples were prepared by solid-state reaction. Powders of La$_{2}$O$_{3}$, Mn$_{2}$O$_{3}$ and NiO were ground altogether with the stoichiometric ratio. The mixed powders were then pressed into half-inch-diameter pellets with 2-3 mm thickness and sintered in flowing air at 1100 $^{\circ}$C for 48 hours.
By means of powder X-ray diffraction, we have confirmed that these samples are without any extra phase except for the BiMnO$_3$ sample, in which unindexed peaks due to some impurity phase were observed but the volume fraction was estimated to be less than 3\%. The \LVO\ sample was confirmed to be a single crystal with use of Laue X-ray diffraction. As for \YTO , we have obtained the clear Laue pattern corresponding to the pseudo-cubic perovskite structure but the orthorhombic $a,b$, and $c$-axis could not be distinguished presumably due to the heavily twined structures.

The magnetization was measured in a Magnetic Property Measurement System (Quantum Design Inc.). The resistivity was measured with use of
Physical Property Measurement System (Quantum Design Inc.). We employ a conventional steady-state method for the measurements of longitudinal and transverse thermal conductivities.  
The longitudinal and transverse temperature gradients $\frac{\partial T}{\partial  x}$ and $\frac{\partial T}{\partial y}$ were measured using both the type E thermal couple ($T \ge $ 20 K) and CX-1050 thermometers ($T \leq$ 50 K). 
The longitudinal thermal conductivity $\kappa_{xx}$ and thermal Hall conductivity $\kappa_{xy}$ were obtained by the following relationships;
\begin{eqnarray}
\kappa_{xx} &=& \frac{\omega_{xx}}{\omega_{xx}^2+\omega_{xy}^2} \approx \frac{1}{\omega_{xx}} \approx -\frac{j_q}{\frac{\partial T}{\partial x}},\\
\kappa_{xy} &=& -\frac{\omega_{xy}}{\omega_{xx}^2+\omega_{xy}^2} \approx -\frac{\omega_{xy}}{\omega_{xx}^2} \approx \frac{\kappa_{xx}^2 \frac{\partial T}{\partial y}}{j_q},
\label{kxy_mes}
\end{eqnarray}
where $\omega_{xx}$, $\omega_{xy}$ and $j_q$ are the longitudinal and Hall components of thermal resistivity and thermal current density, respectively.
While the transverse temperature gradient $\frac{\partial T}{\partial y}$ should be antisymmetric with respect to the magnetic field, the $H$-symmetric component is observed due to the small asymmetry of the thermal probes in the actual measurements as in the transverse voltage of usual electrical Hall measurements. Following the convention of electrical Hall measurement, we have subtracted the $H$-symmetric component using the equation 
$\frac{\partial T}{\partial y}= \frac{\Delta T (+H) -\Delta T (-H)}{2d},$
where $\Delta T$ and $d$ are the observed temperature difference and distance of the transverse thermal probes, respectively. While the Hall conductivity at negative fields obtained by this method is merely the copy of the positive field data, we plot the data at negative fields just for the clarity of figure.
As shown in Eq. (\ref{kxy_mes}), the small value of $\kappa_{xx}$ is desirable in order to estimate the $\kappa_{xy}$ value precisely from the measurement of the transverse temperature gradient $\frac{\partial T}{\partial y}$. The present samples certainly have such small $\kappa_{xx}$ values as discussed later.

\section{Magnon Hall effect in pyrochlore ferromagnets}

\begin{figure}
\begin{center}
\includegraphics*[width=7.6cm]{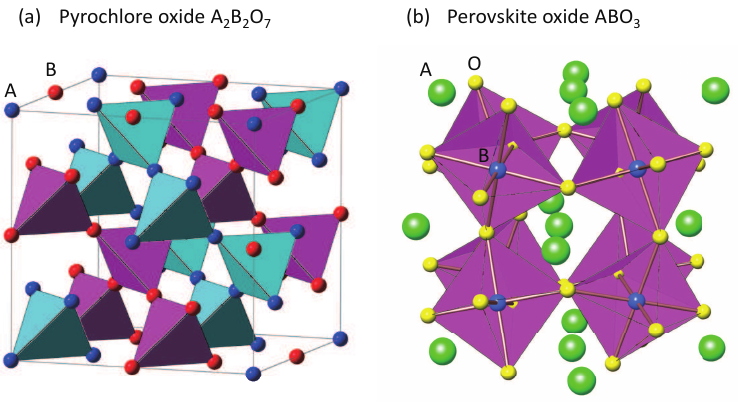}
\caption{(Color online) Crystal structures of (a) pyrochlore oxides A$_2$B$_2$O$_7$ and (b) perovskite oxides ABO$_3$. 
The oxygen ions in (a) are omitted for simplicity.}
\label{fig:1}
\end{center}
\end{figure}

\begin{figure}
\begin{center}
\includegraphics*[width=8.4cm]{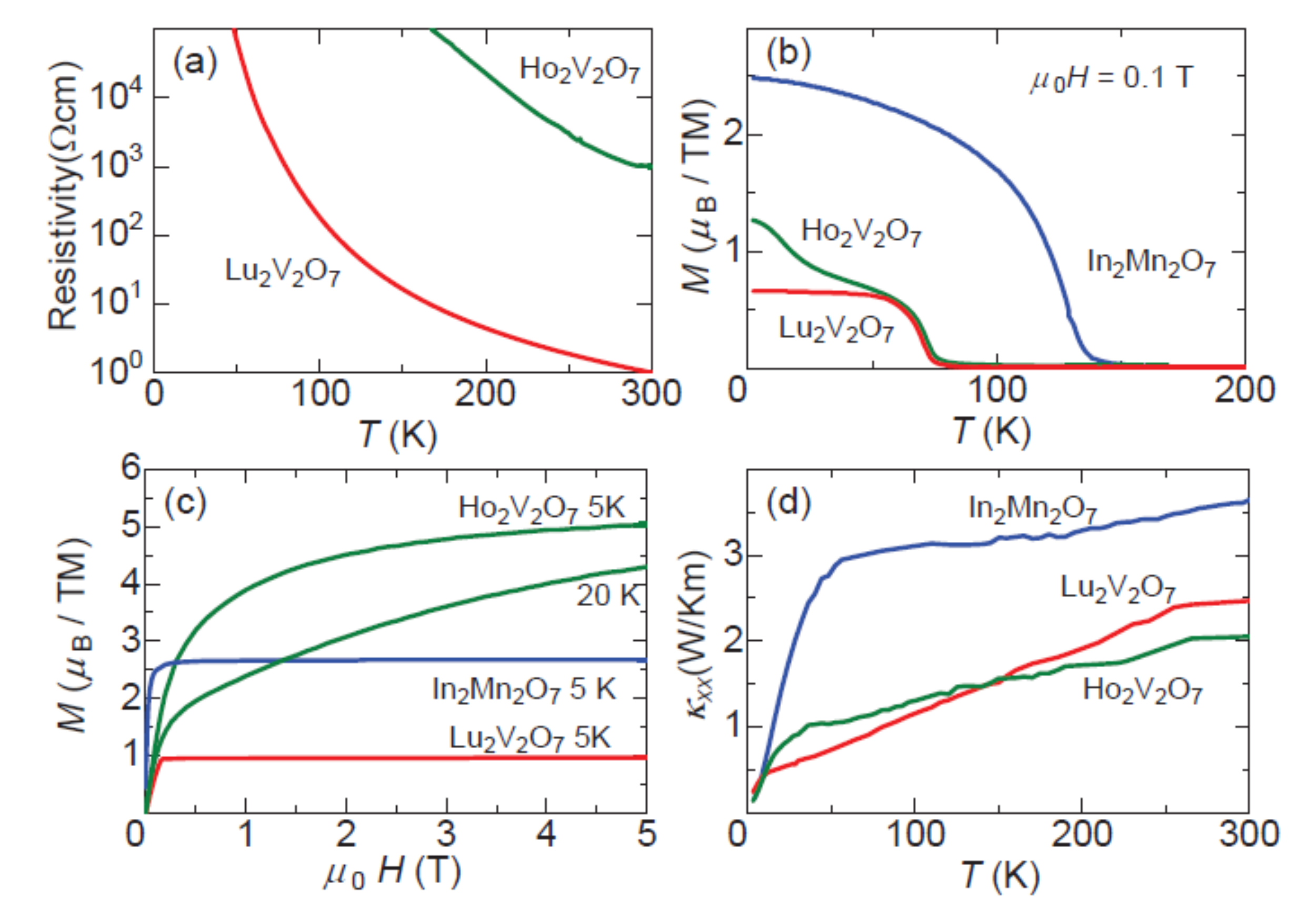}
\caption{(Color online) (a)-(d) Magnetic, electric, and thermal transport properties for Lu$_2$V$_2$O$_7$, Ho$_2$V$_2$O$_7$, and In$_2$Mn$_2$O$_7$. (a)Temperature dependence of the resistivity.
(b) Temperature dependence of the spontaneous magnetization ($\mu_0H = 0.1$ T).
(c)	Magnetization curves at $T$ = 5 K for all the samples. For Ho$_2$V$_2$O$_7$, the 20 K data are also shown. (d)Temperature variation of thermal conductivity.}
\label{fig:2}
\end{center}
\end{figure}

\begin{figure}
\begin{center}
\includegraphics*[width=8.4cm]{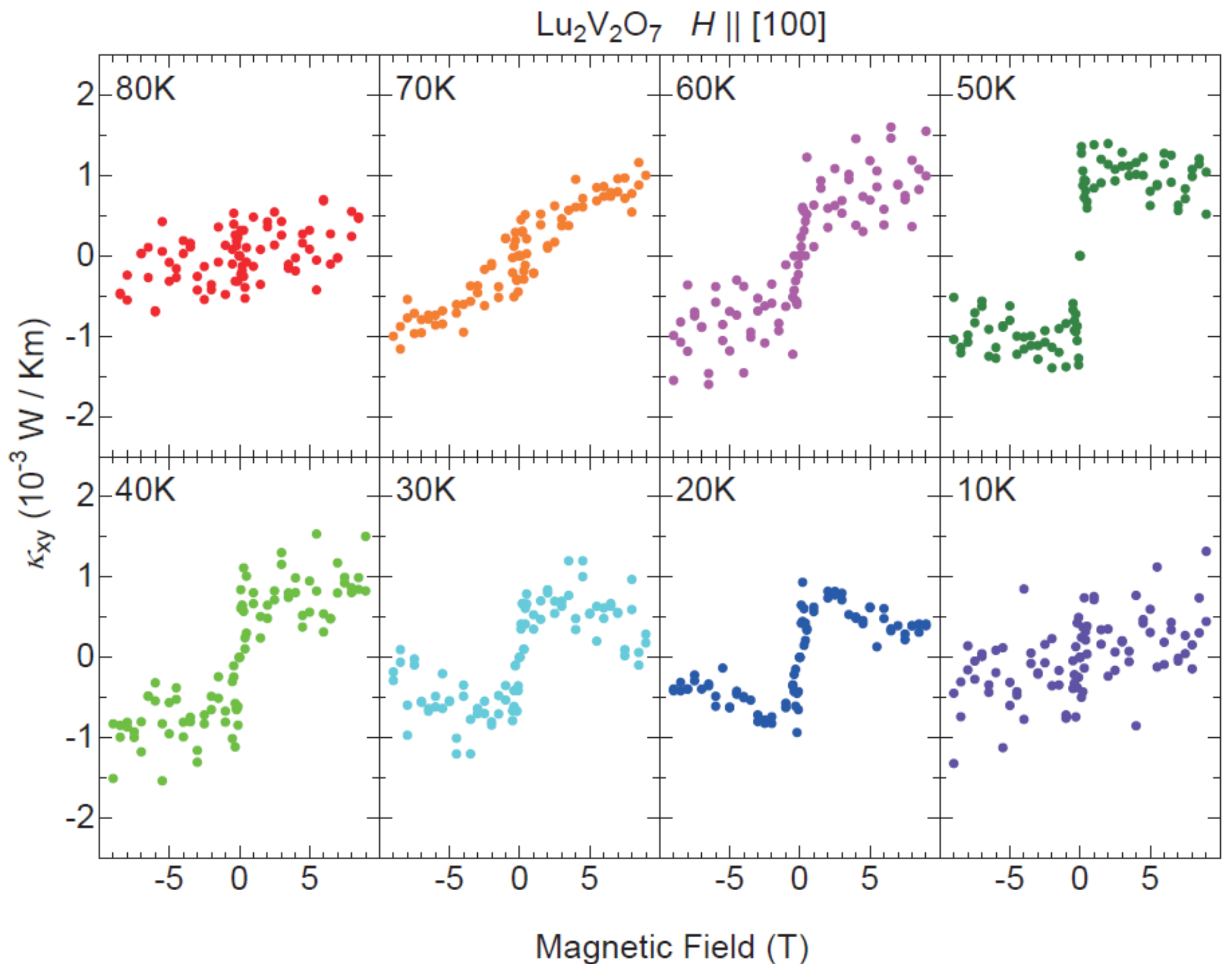}
\caption{(Color online) Thermal Hall conductivity as a function of the magnetic field for \LVO\ at various temperatures. The magnetic field is applied to [100] direction.}
\label{fig:3}
\end{center}
\end{figure}

\begin{figure}
\begin{center}
\includegraphics*[width=8.4cm]{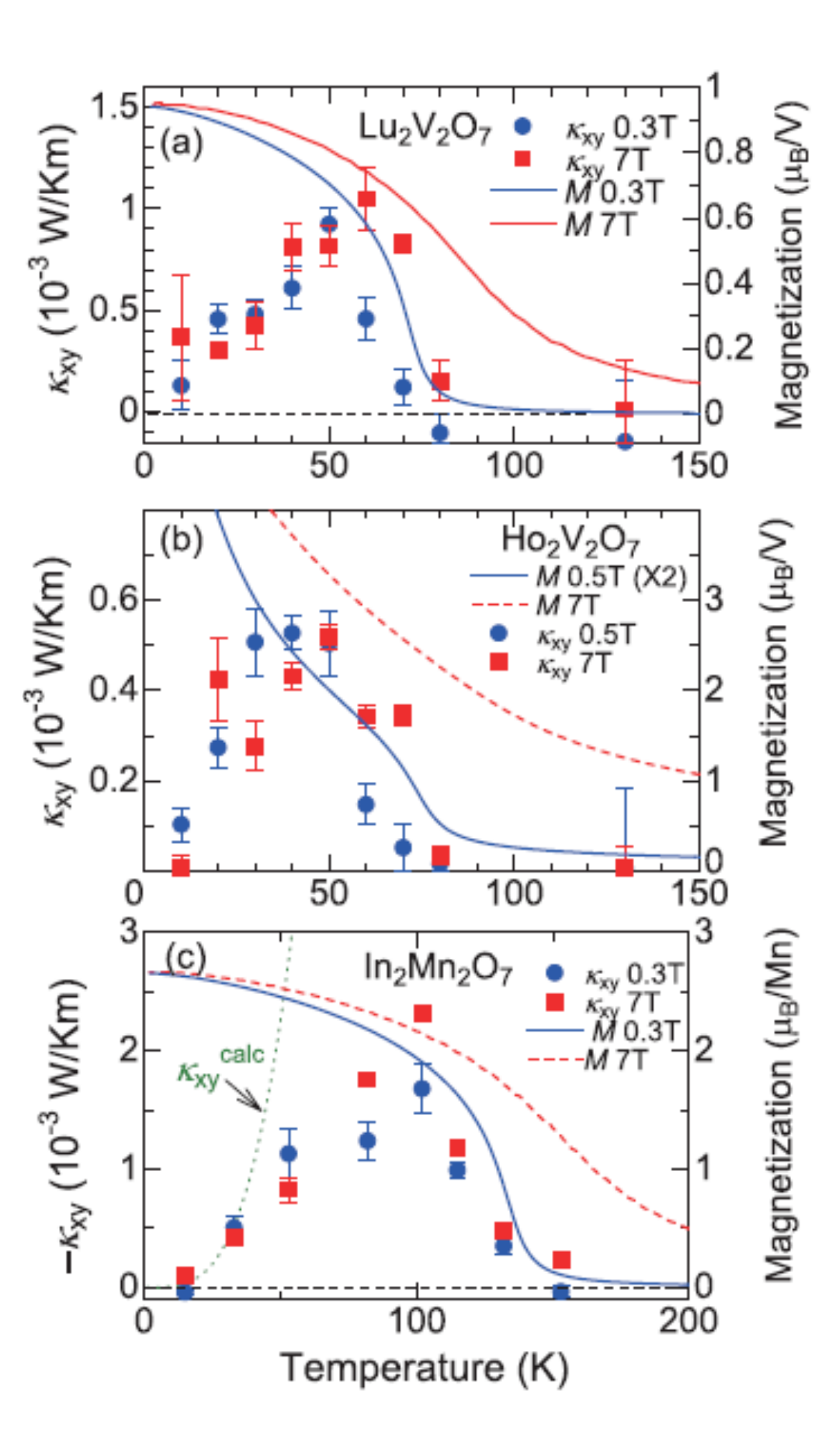}
\caption{(Color online) Temperature dependence of the thermal Hall conductivity and magnetization (a) at $\mu_0 H$ = 0.3 T and $\mu_0 H$ = 7 T for \LVO , (b) at $\mu_0 H$ = 0.5 T and $\mu_0 H$ = 7 T for \HVO , and (c) at $\mu_0 H$ = 0.3 T and $\mu_0 H$ = 7 T for \IMO . $\kappa_{xy}^{\rm calc}$ is the theoretically calculated thermal Hall conductivity for \IMO \  at 0.3 T.}
\label{fig:6}
\end{center}
\end{figure}

\begin{figure}
\begin{center}
\includegraphics*[width=8.4cm]{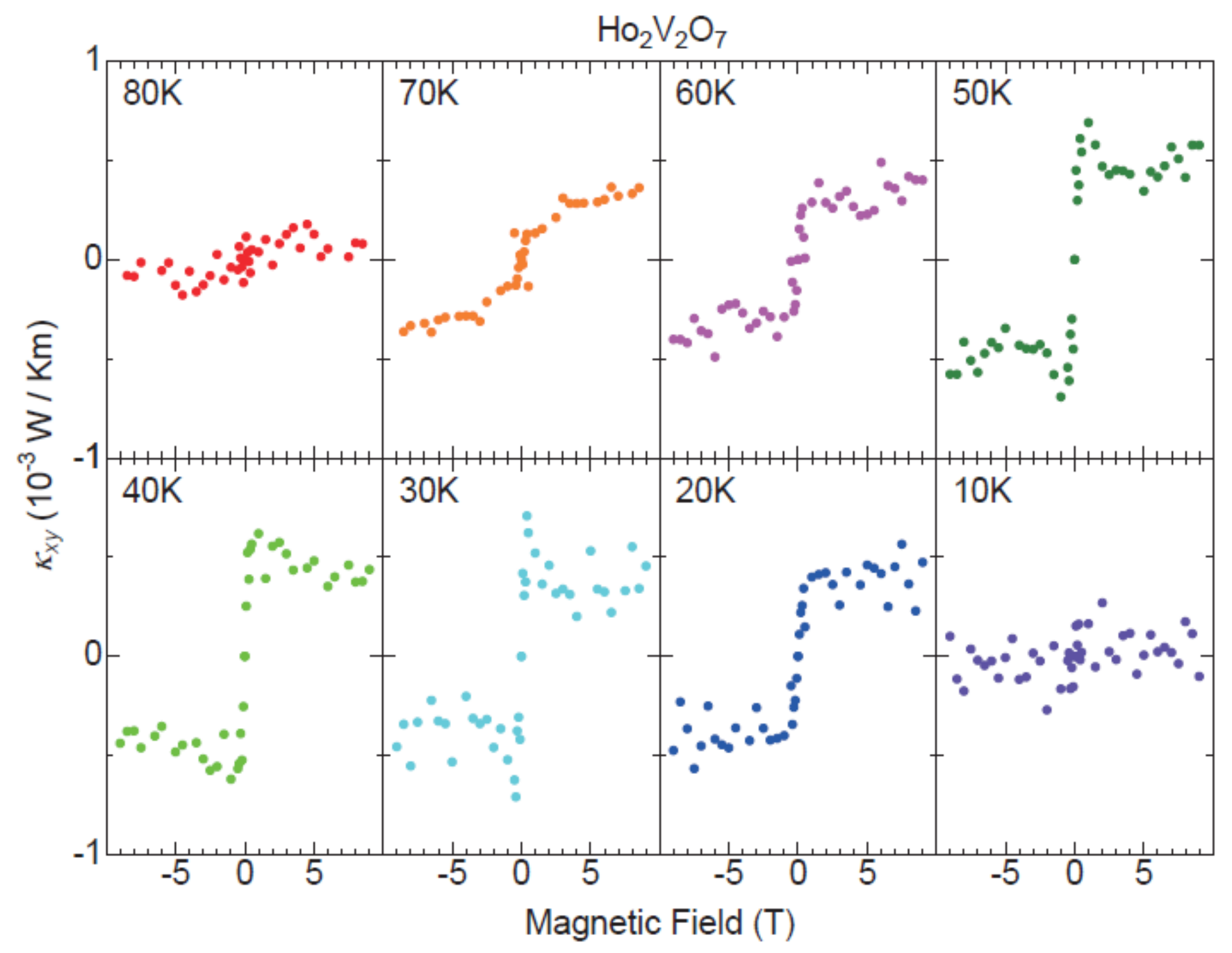}
\caption{(Color online) Magnetic field variation of the thermal Hall conductivity for \HVO\ at various temperatures.}
\label{fig:4}
\end{center}
\end{figure}

\begin{figure}
\begin{center}
\includegraphics*[width=8.4cm]{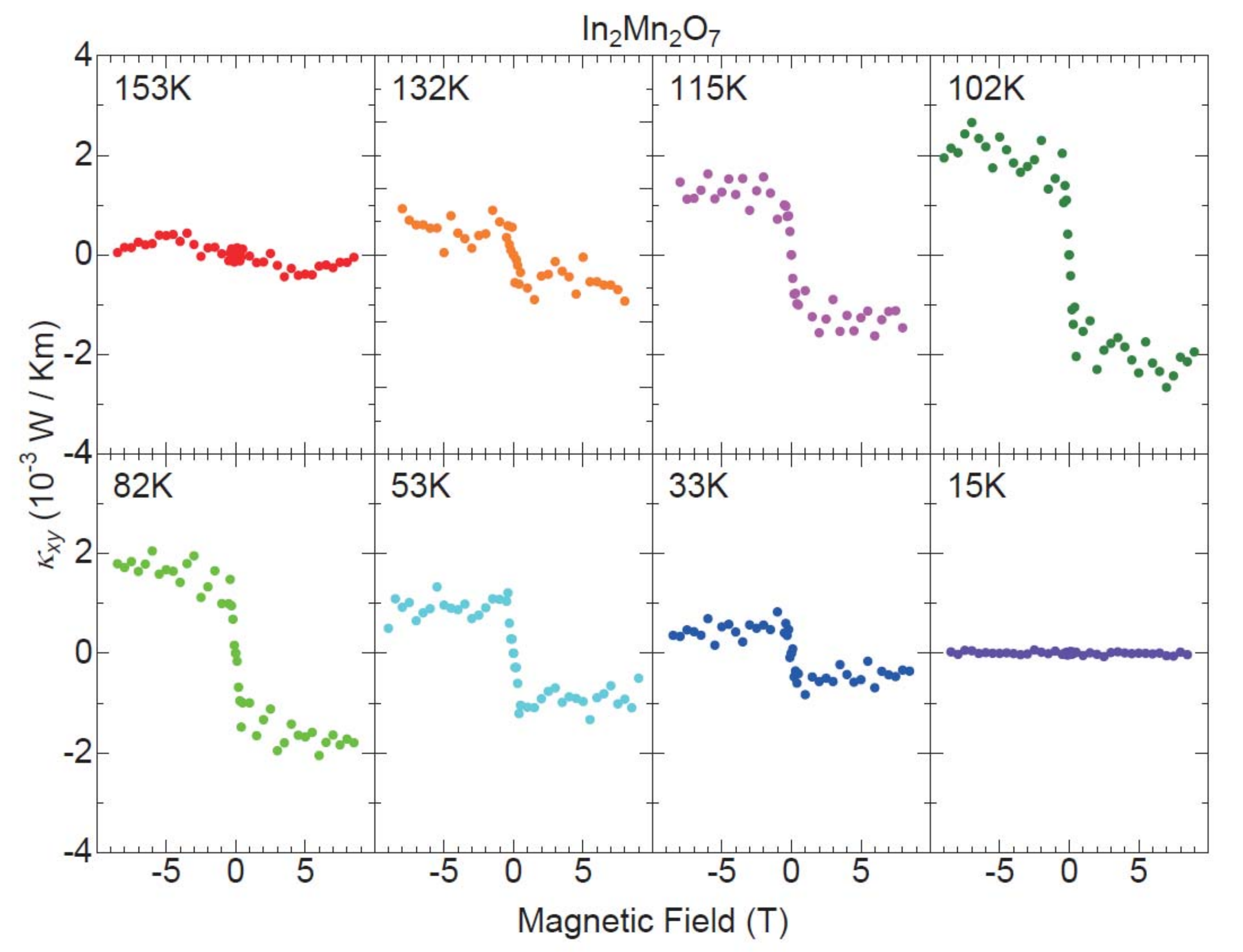}
\caption{(Color online) Magnetic field variation of thermal Hall conductivity for \IMO\ at various temperatures.}
\label{fig:5}
\end{center}
\end{figure}

\begin{figure}
\begin{center}
\includegraphics*[width=8.4cm]{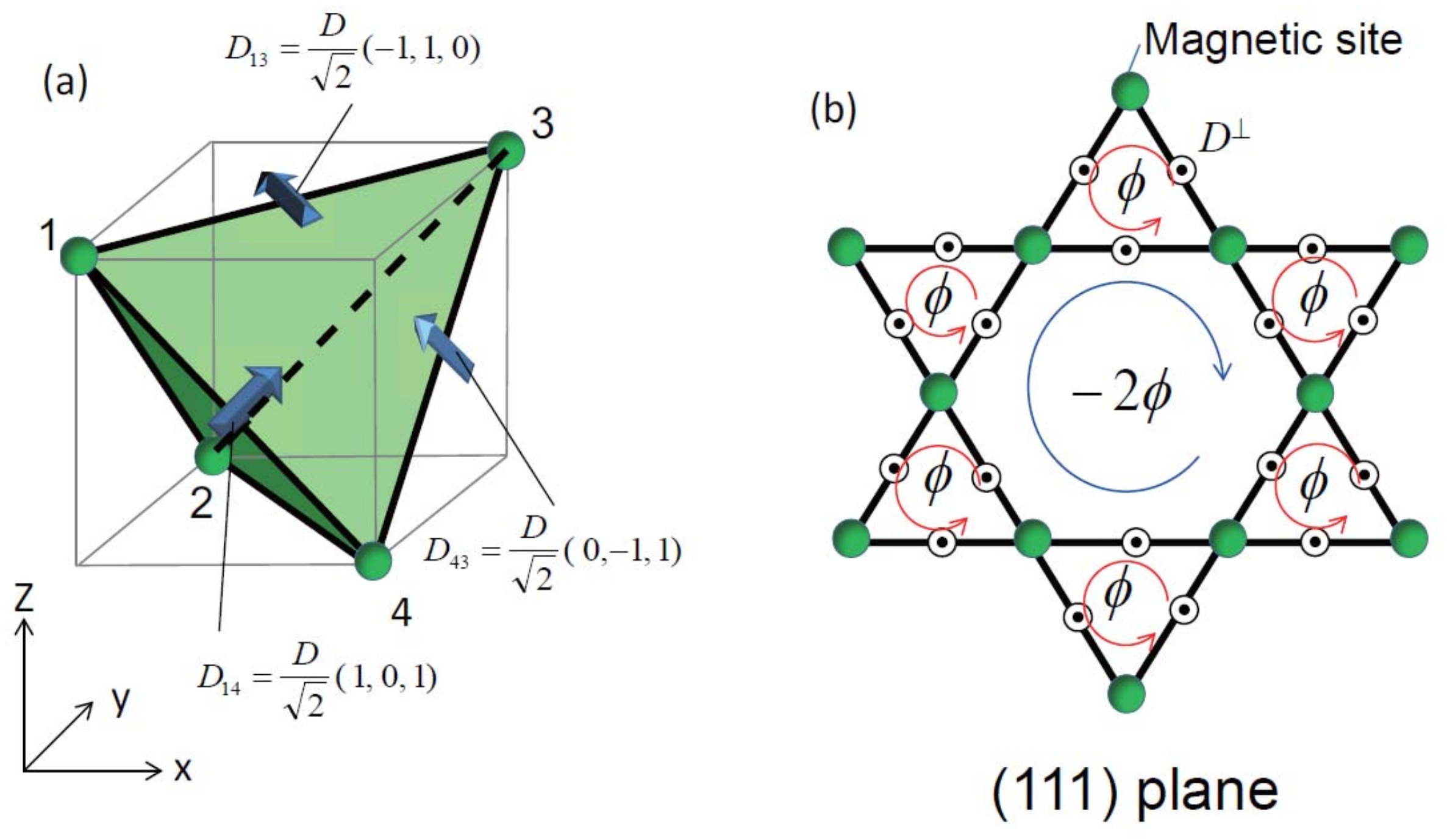}
\caption{(Color online) (a) DM vectors in pyrochlore lattice and (b) magnetic flux 
due to the DM interaction in the (111) plane of the pyrochlore lattice (Kagom$\acute{\rm e}$ lattice).}

\label{fig:7}
\end{center}
\end{figure}

\begin{figure}
\begin{center}
\includegraphics*[width=5cm]{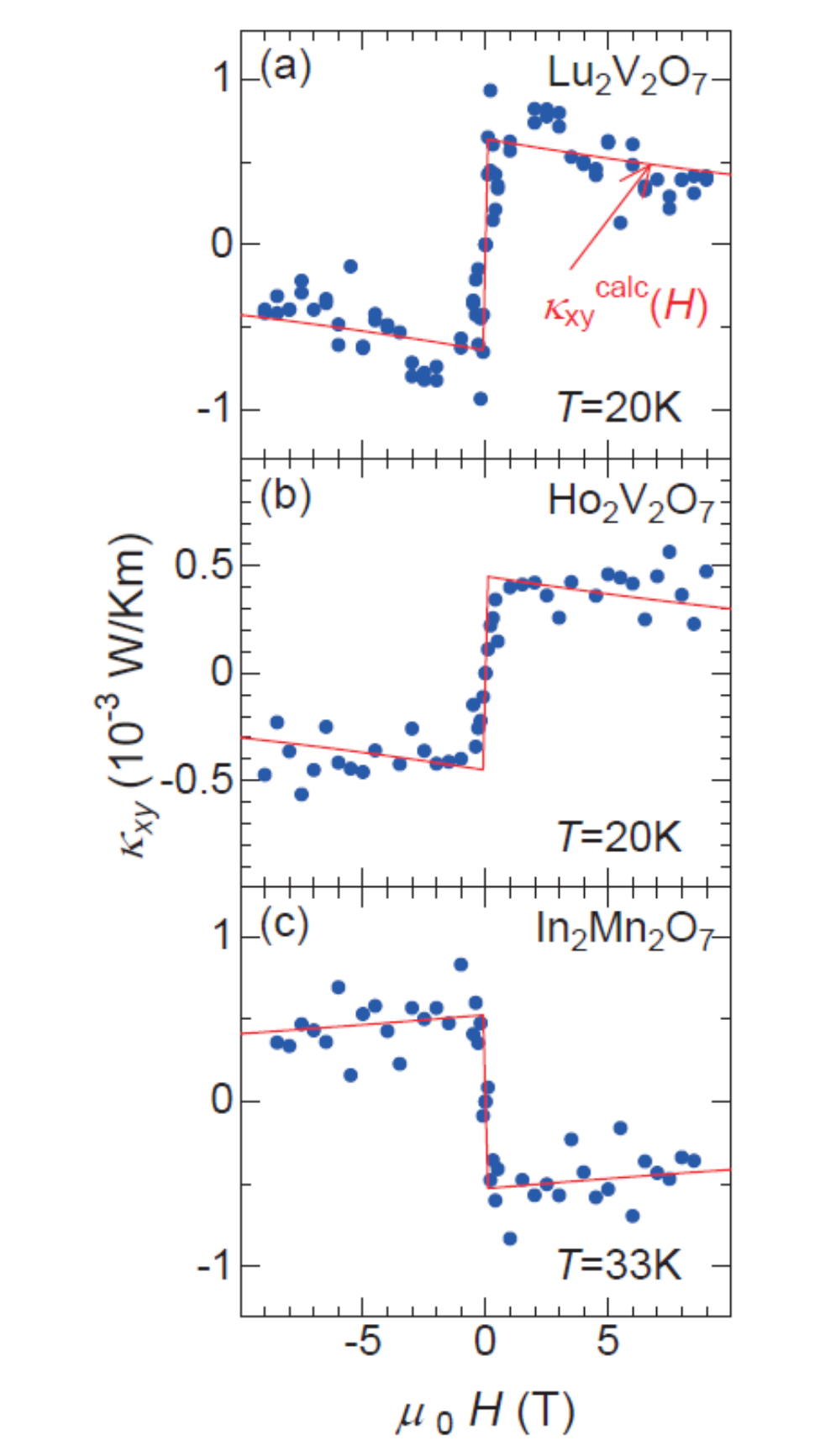}
\caption{(Color online) Magnetic field variation of thermal Hall conductivity (a) at 20 K for \LVO\, (b) at 20 K for \HVO\ and (c) at 33 K for \IMO . The solid line indicates the magnetic field dependence given by the theoretical formula  based on the Dzyaloshinskii-Moriya interaction (Eq. (\ref{eq:kappa_MM})).}
\label{fig:8}
\end{center}
\end{figure}

 In this section, we discuss the thermal Hall conductivity in pyrochlore ferromagnets \LVO , \HVO , and \IMO . The results for Lu$_2$V$_2$O$_7$ have been published in a short paper.\cite{magnon-Hall-exp} Here, the detailed data and analyses for all the three pyrochlore ferromagnets are presented and discussed comprehensively.
Figure~\ref{fig:1}(a) shows the crystal structure of pyrochlore oxide A$_2$B$_2$O$_7$. In this figure, the oxygen ions are omitted for simplicity. The A and B sublattices are identical with each other. The B sublattice is displaced by half a unit cell from the A one. 
The sublattice structure, commonly called pyrochlore lattice, is composed of corner-sharing tetrahedra, and can be viewed as a stacking of alternating Kagome and triangular lattices along the [111] direction.

\LVO\  and \HVO\  are ferromagnetic Mott insulators with one $3d$ electron per vanadium site. 
For these materials, the resistivities increase rapidly with decreasing temperature and the spontaneous magnetization emerges below the Curie temperature \Tc\ $\approx$ 70 K as shown in Figs.~\ref{fig:2}(a) and (b), respectively.  
Spin-polarized neutron diffraction suggests that the $3d$ orbitals are ordered so that they all point to the center of mass of V tetrahedron.\cite{L2V2O7-1} In the orbital ordered state, the virtual hopping process to the higher-lying states stabilizes the ferromagnetic order of $S=1/2$ magnetic moments.\cite{L2V2O7-2}
For Lu$_2$V$_2$O$_7$, the magnetization saturates at low field and the saturated value per V atom nearly coincides with 1 Bohr magneton ($\mu _{\text{B}}$), being consistent with the $d^1$ electronic configuration as shown in Fig.~\ref{fig:2}(c). While only the vanadium ions are magnetic in Lu$_2$V$_2$O$_7$, the Ho ions are also magnetic and behave as Ising spins in Ho$_2$V$_2$O$_7$. Therefore, the saturation field is relatively high and the saturated magnetization with the component of Ho moment is much higher than 1 $\mu_{\rm B}$ in Ho$_2$V$_2$O$_7$.

\IMO\ is also a Mott insulator.\cite{In2Mn2O7-1} The resisivity is too high to be measured even at 300 K. 
There are three $3d$ electrons per manganese site. The localized $S=3/2$ magnetic moments are ferromagnetically coupled to each other due to the strong hybridization among the In $5s$, O~$2p$, and Mn $3d$ orbitals.\cite{In2Mn2O7-2} As shown in Figs.~\ref{fig:2} (b) and (c), the spontaneous magnetization is observed below $T_{\rm C}$=130 K and the saturated magnetic moment nearly coincides with the expected value of 3 $\mu_{\rm B}$.

We show the longitudinal thermal conductivities for \LVO , \HVO , and \IMO\  in Fig. \ref{fig:2}(d).  
According to the Wiedemann-Franz law, the electric contribution of thermal conductivity is less than 10 $^{-4}$ W / Km below 150 K even for the least resistive sample of \LVO , indicating that the heat current is not carried by electronic carriers but by charge neutral excitations such as 
phonons and magnons  
for these samples in this temperature region. The thermal conductivities for these samples are quite small, $2-3$ W/Km even at 300 K and decreases with decreasing temperature. These behaviors are frequently observed in transition-metal oxides, in which both spin and orbital degrees of freedom are active because of the strong electron correlation.\cite{cheng} 
These are likely  because  the mean free paths of phonons and magnons are suppressed due to the fluctuations of spin and orbital in terms of exchange-striction and/or Jahn-Teller coupling. As mentioned above, the small longitudinal thermal conductivity is favorable to the sensitive measurement of $\kappa_{xy}$. 

We reproduce the thermal Hall conductivity for Lu$_2$V$_2$O$_7$ in Fig.~\ref{fig:3}. \cite{magnon-Hall-exp}  
A finite thermal Hall conductivity is observed below $T_{\rm C}$=70 K while it is negligible at 80 K. The magnitude shows a maximum around 50 K and decreases with decreasing temperature from 50 K. As for the magnetic field dependence, the thermal Hall conductivity steeply increases and saturates in the low magnetic field region similarly to the magnetization, which indicates that the observed thermal Hall effect is not normal Hall effect proportional to the magnetic field but the anomalous Hall effect depending on the direction of magnetization.  At low temperatures, the thermal Hall conductivity gradually decreases with magnetic field after the saturation. 

We now discuss the possible carriers of the thermal Hall current, {\it i.e.,} either phonons or magnons.
The thermal Hall effect caused by phonons was previously reported in Tb$_3$Ga$_5$O$_{12}$,\cite{phonon-Hall-exp} and theories based on spin-phonon coupling were proposed to explain it.\cite{phonon-Hall-theory-extrinsic,phonon-Hall-theory-intrinsic} Nevertheless, we have ascribed the thermal Hall conductivity in Lu$_2$V$_2$O$_7$ to the magnon Hall effect for the following reasons. 
Since the mean free path of phonons is expected to increase with magnetic field due to the reduction of spin-phonon scattering, the decrease of the thermal Hall conductivity in the high field region at low temperature cannot be explained in terms of the phonon mechanism. On the other hand, the reduction of the magnon population caused by the gap opening due to Zeeman effect will diminish the magnon contribution of the thermal Hall conductivity in the high magnetic field. In the theory of phonon Hall effect based on the spin-phonon scattering, \cite{phonon-Hall-theory-extrinsic} the thermal Hall angle $\kappa_{xy}$/$\kappa_{xx}$ is 
expected to be proportional to the magnetization. Nevertheless, this is not the case for Lu$_2$V$_2$O$_7$. Figure~\ref{fig:6} (a) shows the temperature dependence of thermal Hall conductivity at $\mu_0 H$=0.3 T and 7 T being compared with the magnetization. 
To be precise, we plot the averaged thermal Hall conductivity between 0.1 T and 0.5 T for the 0.3 T value and between 6.5 T and 7.5 T for the 7 T value because the observed data are scattered. The error bars are estimated by the standard deviation divided by the square root
of the averaged data number.  The thermal Hall conductivity is quite small even at $\mu_0 H$=7 T at $T$=80 K while the magnetization is more than half of the maximum value at this temperature and magnetic field. 
The difference of the temperature dependences in the high magnetic field can be explained  by the magnon picture because the magnon propagation is caused by the ferromagnetic interaction, and is not valid in the magnetic field induced spin polarized state above $T_{\rm C}$. 
Thus, the temperature and magnetic field dependences are well understood with the picture of magnon Hall effect.

In order to further examine the scenario of magnon Hall effect, we have also investigated the thermal Hall conductivity for other pyrochlore ferromagnetic insulators Ho$_2$V$_2$O$_7$ and In$_2$Mn$_2$O$_7$. In Fig.~\ref{fig:4}, we show the magnetic field dependence of thermal Hall conductivity for Ho$_2$V$_2$O$_7$ at various temperatures. The positive thermal Hall conductivity is observed below the transition temperature $T_{\rm C}$=70 K. The magnetic field and temperature dependences are quite similar to the case of Lu$_2$V$_2$O$_7$ while the magnitude is slightly smaller. In Fig.~\ref{fig:6}(b), we show the temperature dependence of thermal Hall conductivity and magnetization at $\mu_0 H$=0.5 T and 7 T. 
(Similarly to the previous case, we plot the averaged thermal Hall conductivity between 0.2 T and 1 T for the 0.5 T value and between 6.5 T and 7.5 T for the 7 T value.) 
The thermal Hall conductivity is quite small above $T_{\rm C}$ even under high magnetic field, similarly to the case of Lu$_2$V$_2$O$_7$. 

The difference from Lu$_2$V$_2$O$_7$ is the presence of Ho $f$ magnetic moment. 
We show the magnetization curve at 20K for Ho$_2$V$_2$O$_7$ in Fig.~\ref{fig:2}(c) to compare with the thermal Hall conductivity.  At this temperature, the magnetization shows a kink structure around 0.3 T but gradually increases with magnetic field even above the kink field. The kink corresponds to the saturation of ferromagnetic vanadium moments and the increase in the high magnetic field is reflected by the gradual alignment process of paramagnetic Ho moments. On the other hand, the thermal Hall conductivity saturates at low magnetic field and does not increase with magnetic field after saturation, which indicates that the thermal Hall effect is caused only by the ferromagnetic and Heisenberg-like vanadium moments. This is consistent with the scenario of magnon Hall effect because the magnon picture is not valid for paramagnetic and Ising-like Ho moments.  

Figure 6 shows the magnetic field dependence of thermal Hall conductivity for In$_2$Mn$_2$O$_7$ at various temperatures. A finite thermal Hall conductivity is observed below $T_{\rm C}$=130 K. While the sign is negative and the magnitude is large (\kxy\ $\approx$ $-2$ $\times$ $10^{-3}$ W/Km at 100 K ) in this case, the temperature and magnetic field dependences are quite similar to the previous cases. The decrease of thermal Hall conductivity after the saturation is also observed at low temperature. 
In Fig.~\ref{fig:6}(c), we show the temperature dependence of thermal Hall conductivity and magnetization at $\mu_0 H$=0.3 T and 7 T. 
(Similarly to the previous cases, we plot the averaged thermal Hall conductivity between 0.2 T and 0.5 T as the 0.3 T value and between 6.5 T and 7.5 T as the 7 T value.) 
The thermal Hall conductivity is fairly suppressed above $T_{\rm C}$ even under magnetic field also in this case. 

 The thermal Hall signal in pyrochlore ferromagnetic insulators can be explained by the theory of magnon Hall effect based on the Dzyaloshinskii-Moriya (DM) interaction.\cite{magnon-Hall-exp} Since the midpoint between any two apices of a tetrahedron is not a center of inversion symmetry in the pyrochlore structure, there is a nonzero DM interaction 
\begin{eqnarray}
H_{\rm DM}=\sum_{\langle i,j \rangle} \DM \cdot (\Spini \times \Spinj ),
\end{eqnarray}
where, $\DM $ and $\Spini $ are, respectively, the DM vector between the sites $i$ and $j$, and the TM spin moment at the site $i$. The sum is taken over all pairs of neighboring sites. 
As shown in Fig.~\ref{fig:7}(a), the DM vectors on a single tetrahedron of the pyrochlore lattice are determined by Moriya's rule \cite{Moriya, Elhajal, Kotov} and distributed as
\begin{eqnarray}
\D _{13}&=& \frac{D}{\sqrt{2}}(-1,1,0),
\D _{24}= \frac{D}{\sqrt{2}}(-1,-1,0),\\
\D _{43}&=& \frac{D}{\sqrt{2}}(0,-1,1),
\D _{12}= \frac{D}{\sqrt{2}}(0,-1,-1),\\
\D _{14}&=& \frac{D}{\sqrt{2}}(1,0,1),
\D _{23}= \frac{D}{\sqrt{2}}(1,0,-1),
\end{eqnarray}
where 1,2,3 and 4 denote the sites shown in Fig. \ref{fig:7} (a). 
The DM vector on each bond is perpendicular to the bond and parallel to the face of the surrounding cube. 
Let us see that the DM interaction does not disturb the ferromagnetic ordering.  
To this end, we replace $\Spini$ with $(\langle \bm{S} \rangle +\delta \Spini )$, 
where $\langle \bm{S} \rangle$ denotes the ordered moment.  
For a single tetrahedron, the DM interaction gives rise to the first order term in the fluctuation $\delta \Spini $ as
\begin{eqnarray}
\delta H^{\boxtimes}_{\text{DM}}=\sum^{4}_{i=1}\sum _{j(\neq i)}\bm {D}_{ij}\cdot (\delta \bm {S}_{i} \times \langle \bm {S}\rangle),
\end{eqnarray}
where the superscript $\boxtimes$ denotes the single tetrahedron shown in Fig. \ref{fig:7} (a). 
However, since $\sum _{j(\neq i)}\DM =0$ for any $i$, one can confirm $\delta H_{\text{DM}}=0$. This implies that the collinear ferromagnetic ground state is stable against the DM interaction.

We now briefly explain how the DM interaction can act on the magnons like an effective gauge field 
and gives rise to the Berry phase effect. 
The Bloch state of a single magnon with momentum $\bm{k}$ is defined by 
\begin{eqnarray}
|\bm {k}\rangle \equiv \frac{1}{\sqrt{N}}\sum _{i}e^{i\bm {k}\cdot\bm {R}_i}| i \rangle,
\end{eqnarray}
where $|i\rangle$ is the magnon state, in which the spin state at the position $\bm{R}_i$ is $S-1$ 
with all the other spins being completely aligned with the $H$ direction. 
The matrix element corresponding to the transfer integral of magnons is expressed as 
\begin{eqnarray}
&& \langle i|-J\bm {S}_{i}\cdot \bm {S}_{j}+\bm {D}_{ij}\cdot (\bm {S}_{i}\times \bm {S}_{j})|j \rangle \nonumber \\
&=& -\frac{1}{2}\langle i|\Tilde{J}(e^{-i\phi}S_{i}^{+}S_{j}^{-}+e^{i\phi}S_{i}^{-}S_{j}^{+})|j \rangle =-\frac{\Tilde{J}}{2}e^{i\phi},
\label{eq:matrix_element}
\end{eqnarray}
where $J$ is the nearest-neighbor ferromagnetic exchange, and 
$S^{\pm}$ is the operator that increases (decreases) the spin component along the direction $\bm {n}=\bm {H}/|\bm {H}|$. 
The relation 
$\Tilde{J}e^{i\phi}=J+i \bm{D}_{ij} \cdot \bm {n}$ determines the complex transfer integral for magnons. 
Note that the component of the DM vector perpendicular to $\bm{n}$ does not contribute to 
the spin-wave Hamiltonian up to quadratic order. 
The phase factor due to the DM interaction can be viewed as a \lq\lq fictitious magnetic flux\rq\rq. 
It should be noted that
the lattice geometry is important to avoid cancellation of the effect of phase factor; the inequivalent loops in the unit cell are necessary for a finite thermal Hall conductivity as suggested previously.\cite{magnon-Hall-theory-2}
There are such inequivalent loops in the unit cell in the pyrochlore lattice. Figure \ref{fig:7}(b) shows the (111) plane cross-section of pyrochlore lattice, which is the Kagom$\acute{\rm e}$ lattice. The out-of-plane component of DM vector perpendicular to the (111) plane is also depicted in this figure. The Kagom$\acute{\rm e}$ lattice is certainly composed of inequivalent loops, namely the triangles and hexagons. While the total magnetic flux in the unit cell is zero, 
the Berry curvature, $i.e.$ the fictitious magnetic flux in momentum space, becomes nonzero  
owing to the inequivalence of the loops, which may induce the thermal Hall effect of magnons.  

We now turn to the quantitative calculation of the thermal Hall effect 
and its comparison with the experimental results.  
A formula for the thermal Hall conductivity of magnon systems was first proposed in Refs. \onlinecite{magnon-Hall-theory-2, magnon-Hall-exp}. 
However, it has recently pointed out by Matsumoto and Murakami that 
an additional term corresponding to the rotational motions of magnons is missing in this formula.\cite{Matsumoto-Murakami}
In the following, we will first present the derivation of 
the previous formula~\cite{magnon-Hall-theory-2, magnon-Hall-exp} and show how this formula 
relates the thermal Hall conductivity to the Berry curvature in momentum space. 
Then we will present the correct formula based on the new formulation, 
which is also expressed in terms of the Berry curvature.\cite{Matsumoto-Murakami} 
The ratio of the DM interaction to the ferromagnetic exchange interaction $D/J$ 
will also be estimated using the new formula.

Let us first provide a brief synopsis of the derivation of the previous formula. 
We start from the spin Hamiltonian consisting of the ferromagnetic exchange interactions, 
the DM, and Zeeman terms:  
\begin{eqnarray}
H &=& H_0 + H_{\rm DM},
\\
H_0 &=& -J\sum_{\langle i,j \rangle} {\bm S}_i \cdot {\bm S}_j - g \mu_{\rm B} {\bm H} \cdot \sum_i {\bm S}_i.  
\end{eqnarray}
We can derive the spin-wave Hamiltonian 
appying the Holstein-Primakoff transformation, which is a slight modification of 
the calculation of the matrix element Eq. (\ref{eq:matrix_element}) presented above. 
The spin-wave Hamiltonian in momentum space reads
\begin{equation}
H_{\rm SW} =
\sum_{\bm k} \Psi^\dagger_{\bm k} {\cal H} ({\bm k})
\Psi_{\bm k},
\end{equation}
where $\Psi ({\bm k}) = (b_{1,{\bm k}}, b_{2, {\bm k}}, b_{3,{\bm k}}, b_{4,{\bm k}})^{\rm T}$ 
are the boson operators that annihilate magnons, and 
\begin{widetext}
\begin{eqnarray}
{\cal H}({\bm k}) =-2JS 
\left(\begin{array}{cccc}  
0  & e^{i\phi_{12}} \frac{\cos(k_y-k_z)}{\cos(\phi_{12})} &  e^{i\phi_{13}} \frac{\cos(k_x+k_y)}{\cos(\phi_{13})} & e^{i\phi_{14}} \frac{\cos(k_z-k_x)}{\cos (\phi_{14})} \\
 & 0 & e^{i\phi_{23}} \frac{\cos(k_x+k_z)}{\cos(\phi_{23})} & e^{i\phi_{24}} \frac{\cos(k_x-k_y)}{\cos(\phi_{24})} \\
 &  & 0 &  e^{-i\phi_{43}} \frac{\cos(k_y+k_z)}{\cos(\phi_{43})}\\
 &  &  & 0
\end{array}\right) + {\rm const.}, 
\label{eq:sHam_pyro}
\end{eqnarray}
\end{widetext}
where $k_\alpha = {\bm k}\cdot {\hat \alpha}$ ($\alpha = x,y,z$).  
The lower triangle of the matrix is understood to be filled so that the matrix is hermitian.  
The phase factors $\phi_{ij}$ in Eq. (\ref{eq:sHam_pyro}) are uniquely determined by 
the DM vectors and the direction of the magnetic field ${\bm n}=(n_x, n_y, n_z)$ 
through the relation $\tan \phi_{ij} = ({\bm D}_{ij}\cdot {\bm n})/J$ 
(see Supporting Online Material of Ref.~\onlinecite{magnon-Hall-exp} for details). 

One can diagonalize the spin-wave Hamiltonian at each ${\bm k}$ and obtain the eigenvector 
$|\psi_m ({\bm k})\rangle$ with corresponding eigenvalues $\omega_m ({\bm k})$, 
where $m$ ($m=1,2,3,4$) is the band index.  
Using the Kubo formula, one can calculate the thermal Hall conductivity $\kappa_{\alpha \beta}$. 
In the low-temperature region, the dominant contribution  comes from the small momentum region of 
the lowest ($m=1$) magnon band owing to the Bose nature of magnons 
and the fact that the lowest band is well separated from the other bands near ${\bm k}={\bm 0}$. 
Retaining only the first-order terms in the DM interaction, the analytic expression for the anomalous thermal Hall conductivity due to magnons is obtained as
\begin{eqnarray}
{\bar \kappa}_{\alpha \beta} &\approx &
-\frac{\Delta^2}{2T} \int_{\text{BZ}}\frac{d^{3}k}{(2\pi)^{3}} \rho_1 ({\bm k})\,
{\rm Im} \bigg\langle \frac{\partial \psi_{1}(\bm {k})}{\partial k_{\alpha}}
\bigg| \frac{\partial \psi_{1}(\bm {k})}{\partial k_{\beta}} \bigg\rangle \nonumber\\
&\approx & \Phi _{\alpha \beta}\frac{k_{\text{B}}^{2}T}{\pi ^{3/2}\hbar a}(2+\frac{g\mu _{B}H}{2JS})^{2}\sqrt{\frac{k_{\text{B}}T}{2JS}}\text{Li}_{\frac{5}{2}}(e^{-\frac{g\mu_{\rm B}H}{k_{\text{B}}T}}), ~~~~\,
\label{eq:kxy}
\end{eqnarray}
where $\Delta=8JS+2g \mu_{\rm B}H$, 
$\rho_m ({\bm k})=[\exp (\beta \omega_m ({\bm k}))-1]^{-1}$ denotes the Bose distribution function, 
and the integral is over the Brillouin zone (BZ).  
$\psi_1(\bm{k})$ and $\omega_1({\bm k})$ are, respectively, the eigenvector of ${\cal H}({\bm k})$ corresponding to 
the lowest magnon mode and its dispersion. 
In the second line of Eq. (\ref{eq:kxy}), 
$\text{Li}_{n}(z)=\sum^{\infty}_{k=1}\frac{z^{k}}{k^{n}}$ is the polylogarithm, 
$a$  the lattice constant, and $\Phi _{\alpha \beta}=\epsilon _{\alpha \beta \gamma}n_{\gamma}D/(8\sqrt{2}J)$ 
with the totally antisymmetric tensor $\epsilon _{\alpha \beta \gamma}$.  
We have thus obtained the expression for $\kappa_{xy}$ in terms of the Berry curvature 
in the first line of Eq. (\ref{eq:kxy}) as a result of the approximation neglecting the upper bands. 
However, as recently pointed out by Matsumoto and Murakami~\cite{Matsumoto-Murakami}, 
the thermal Hall conductivity of magnons is, {\it in general}, expressed by the Berry curvature 
without such an approximation. 
By noting that there is an additional correction to Eq. (\ref{eq:kxy}) that corresponds to 
the rotational motions of magons, they derived the following formula:
\begin{equation}
\kappa_{\alpha \beta} = 2T \sum_n \int_{\text{BZ}}\frac{d^{3}k}{(2\pi)^{3}} c_2 ( \rho_n({\bm k}))\,
{\rm Im}\bigg\langle \frac{\partial \psi_{n}(\bm {k})}{\partial k_{\alpha}}
\bigg| \frac{\partial \psi_{n}(\bm {k})}{\partial k_{\beta}} \bigg\rangle
\end{equation}
where $c_2 (\rho) = (1+\rho) \left( \log \frac{1+\rho}{\rho} \right)^2 - (\log \rho)^2 -2 {\rm Li}_2 (-\rho)$. 
Using the approximation that neglects the upper bands, we have
\begin{eqnarray}
\kappa_{\alpha \beta} &\approx & 2T \int_{\text{BZ}}\frac{d^{3}k}{(2\pi)^{3}} c_2 ( \rho_1 ({\bm k}))\,
{\rm Im}\bigg\langle \frac{\partial \psi_{1}(\bm {k})}{\partial k_{\alpha}}
\bigg| \frac{\partial \psi_{1}(\bm {k})}{\partial k_{\beta}} \bigg\rangle
\nonumber \\
&=& \Phi_{\alpha \beta} \frac{4 k^2_{\rm B}T}{3 \pi^2 \hbar a} \left( \frac{k_{\rm B}T}{2 JS} \right)^{5/2}
\nonumber \\
&& 
\times \int^\infty_0 c_2 \left( \frac{1}{e^{t+\frac{g \mu_{\rm B}H}{k_{\rm B}T}}-1}  \right) t^{3/2} dt.
\label{eq:kappa_MM}
\end{eqnarray}
Here we have restored the unit $\hbar=k_{\rm B}=a/4=1$. 
Note that the third line of Eq. (\ref{eq:kappa_MM}) is a dimensionless integral. 
The temperature and field dependencies of $\kappa_{\alpha \beta}$ are clearly different from those of ${\bar \kappa}_{\alpha \beta}$ 
in Eq. (\ref{eq:kxy}). 

 In Figs. 8 (a)-(c), we show the fitting of the thermal Hall conductivity at low temperature by Eq. (\ref{eq:kappa_MM}) for the pyrochlore ferromagnetic insulators. 
The ferromagnetic exchange interaction  $J$ is estimated by the specific heat data\cite{magnon-Hall-exp} for Lu$_2$V$_2$O$_7$ and the mean field values $J=k_{\text{B}}T_{\text{C}}/4S(S+1)$ are employed for Ho$_2$V$_2$O$_7$ and In$_2$Mn$_2$O$_7$.
The field dependence is well reproduced by Eq. (\ref{eq:kappa_MM}) as shown in these figures. From the fitting, we estimate the ratio of the DM interaction to the ferromagnetic exchange interaction $D/J$ as $D/J=-0.38$ for \LVO , $D/J=-0.07$ for \HVO\ and $D/J=0.018$ for \IMO . It should be noted that the estimated $D/J$ values have large error bars because it is quite sensitive to the estimate of $J$. 
(We have the estimate for $|D/J| \sim  0.007-0.035$ in the case of In$_2$Mn$_2$O$_7$ if the error in the estimate of $J$ is within 30 percent.)
Nevertheless, the estimated order of magnitude $|D/J| \sim 10^{-1}-10^{-2}$ is still meaningful, which is comparable to those reported in TM oxides.\cite{CdCr2O4,LaMnO3} 
The difference of the sign of $D/J$ between In$_2$Mn$_2$O$_7$ and the pyrochlore vanadates may be ascribed to the different electronic configuration ($d^1$ for Lu$_2$V$_2$O$_7$ and Ho$_2$V$_2$O$_7$, and $d^3$ for In$_2$Mn$_2$O$_7$) while the accurate estimation of many virtual hopping processes is required to determine $D$ and $J$.
The first-principle band structure calculation is the most reliable approach to theoretically estimate the $D/J$ as employed by Xiang {\it et al.} for Y$_2$V$_2$O$_7$ using DFT+U calculation.\cite{bandcalcRVO} The detailed comparison of the Dzyaloshinskii-Moriya interaction between pyrochlore vanadates and In$_2$Mn$_2$O$_7$ with band calculation is a future problem.
As exemplified in Fig. 4(c), theoretically calculated thermal Hall conductivity $\kappa_{xy}^{\rm calc}$ steeply increases with temperature. This is qualitatively consistent with the experimental observation in the low temperature region. The thermal Hall signal is within the noise level at 10 K but clearly observed around 20 K-30 K for all the pyrochlore ferromagnets as shown in Figs. 3, 5, and 6. These may be caused by the steep temperature dependence of magnon Hall effect. The weaker increase of experimentally observed thermal Hall conductivity in the higher temperature region seems to be caused by the effect of magnon-magnon interaction, which is important in this temperature range but neglected in the theory.

As we have seen, the thermal Hall signal due to the magnon Hall effect is commonly observed in the pyrochlore ferromagnetic insulators. 
The order of magnitude and the temperature/field dependences are quite similar among the materials. The phenomenon of magnon Hall effect seems generic in pyrochlore ferromagnets.

\section{Magnon Hall effect in perovskite lattice systems}

\begin{figure}
\begin{center}
\includegraphics*[width=8.4cm]{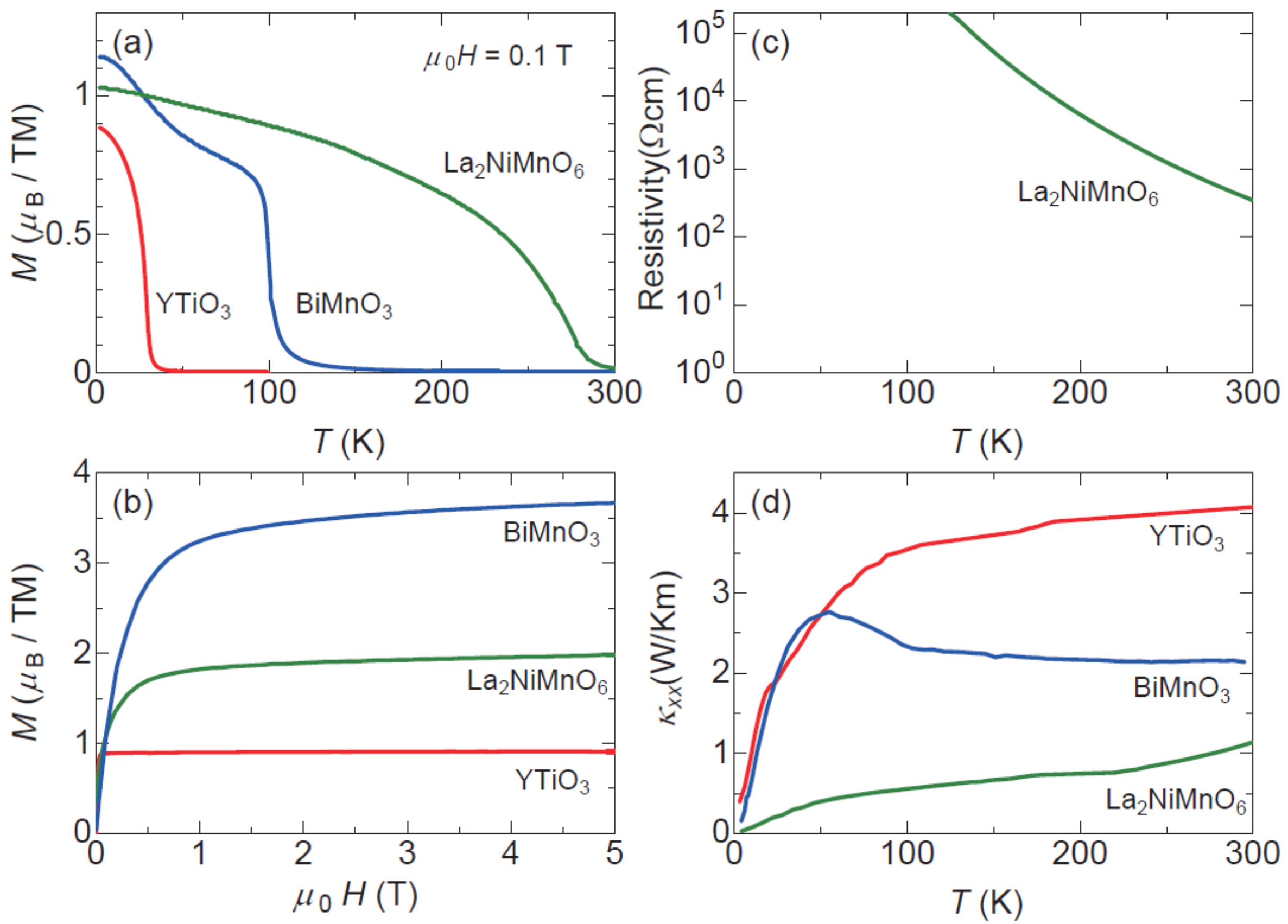}
\caption{(Color online). (a)-(d) Magnetic, electric, and thermal properties of ferromagnetic perovskite oxides La$_2$NiMnO$_6$, BiMnO$_3$, and YTiO$_3$.
(a)	Temperature dependence of magnetization at a magnetic field $\mu_0 H$ = 0.1 T.
(b)	Magnetization curves at $T$ = 5 K.
(c)	Temperature variation of resistivity.
(d)	Temperature variation of longitudinal  thermal conductivity.}
\label{fig:9}
\end{center}
\end{figure}

\begin{figure}
\begin{center}
\includegraphics*[width=6cm]{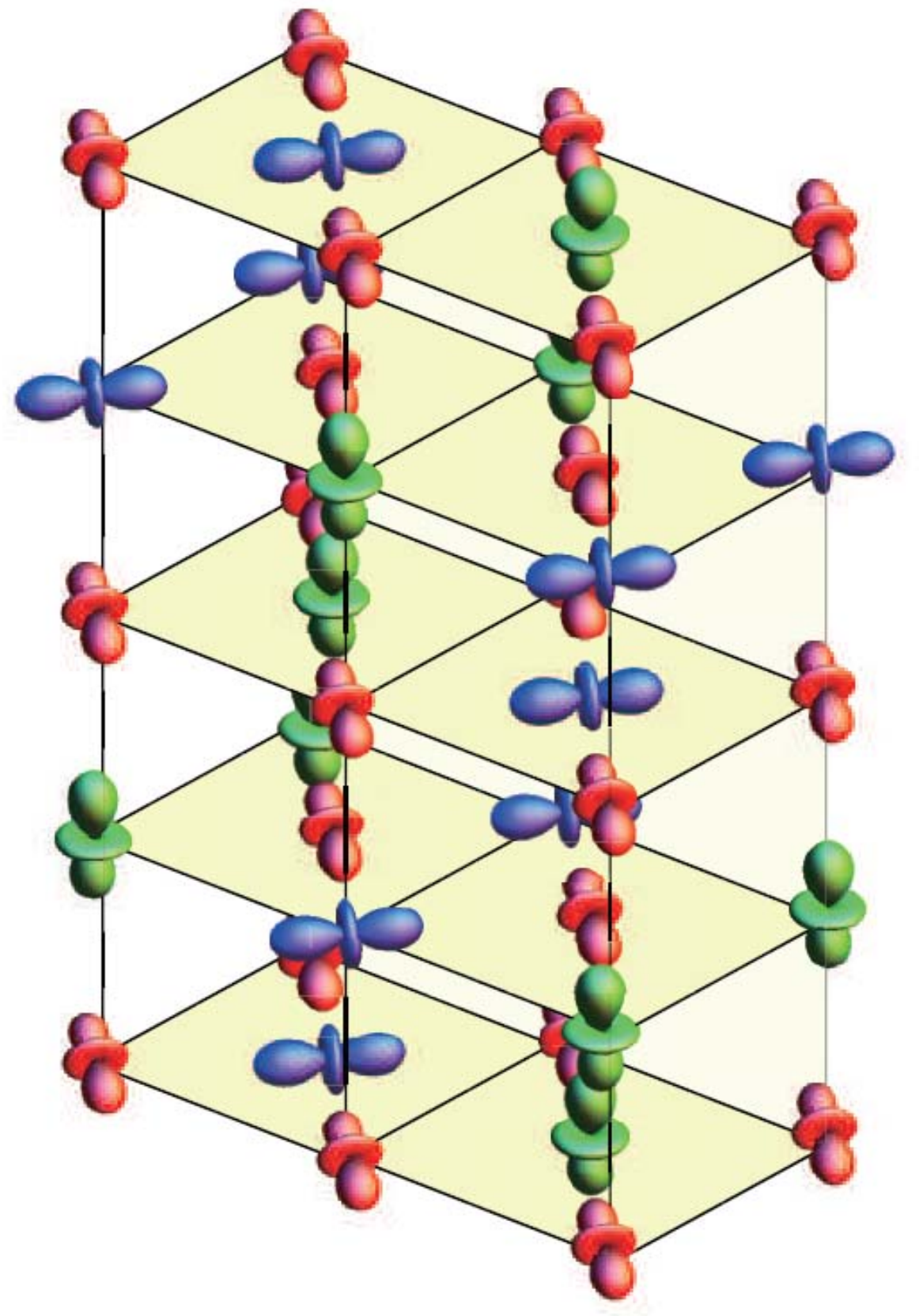}
\caption{(Color online). Orbital ordering in \BMO.}
\label{fig:10}
\end{center}
\end{figure}

\begin{figure}
\begin{center}
\includegraphics*[width=8.4cm]{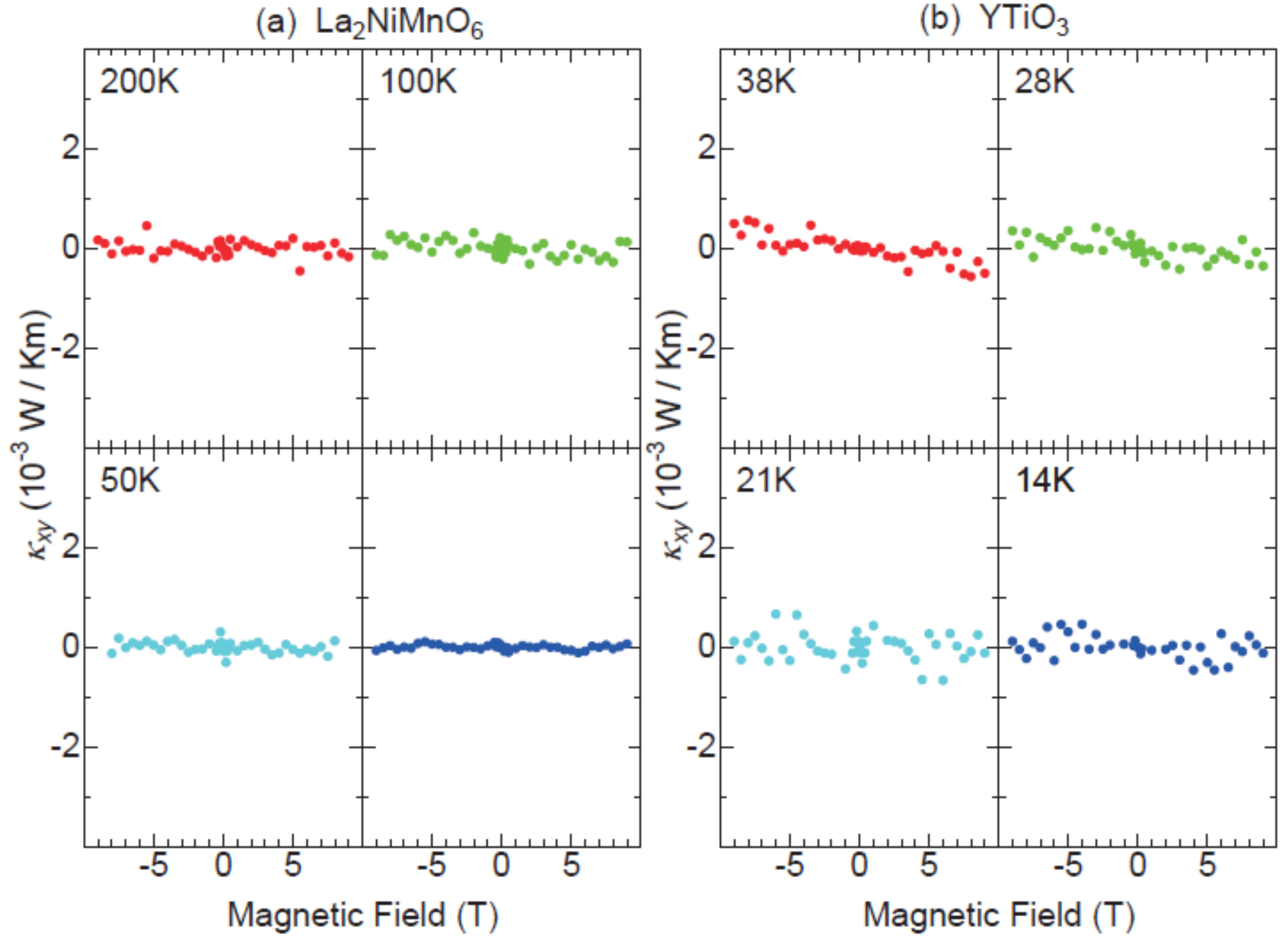}
\caption{(Color online). (a),(b) Magnetic field variation of the thermal Hall conductivity for (a)\LNMO\ and (b)\YTO.  For \YTO , magnetic field is applied to the [100] direction of pseudocubic crystal.}
\label{fig:11}
\end{center}
\end{figure}

\begin{figure}
\begin{center}
\includegraphics*[width=8.4cm]{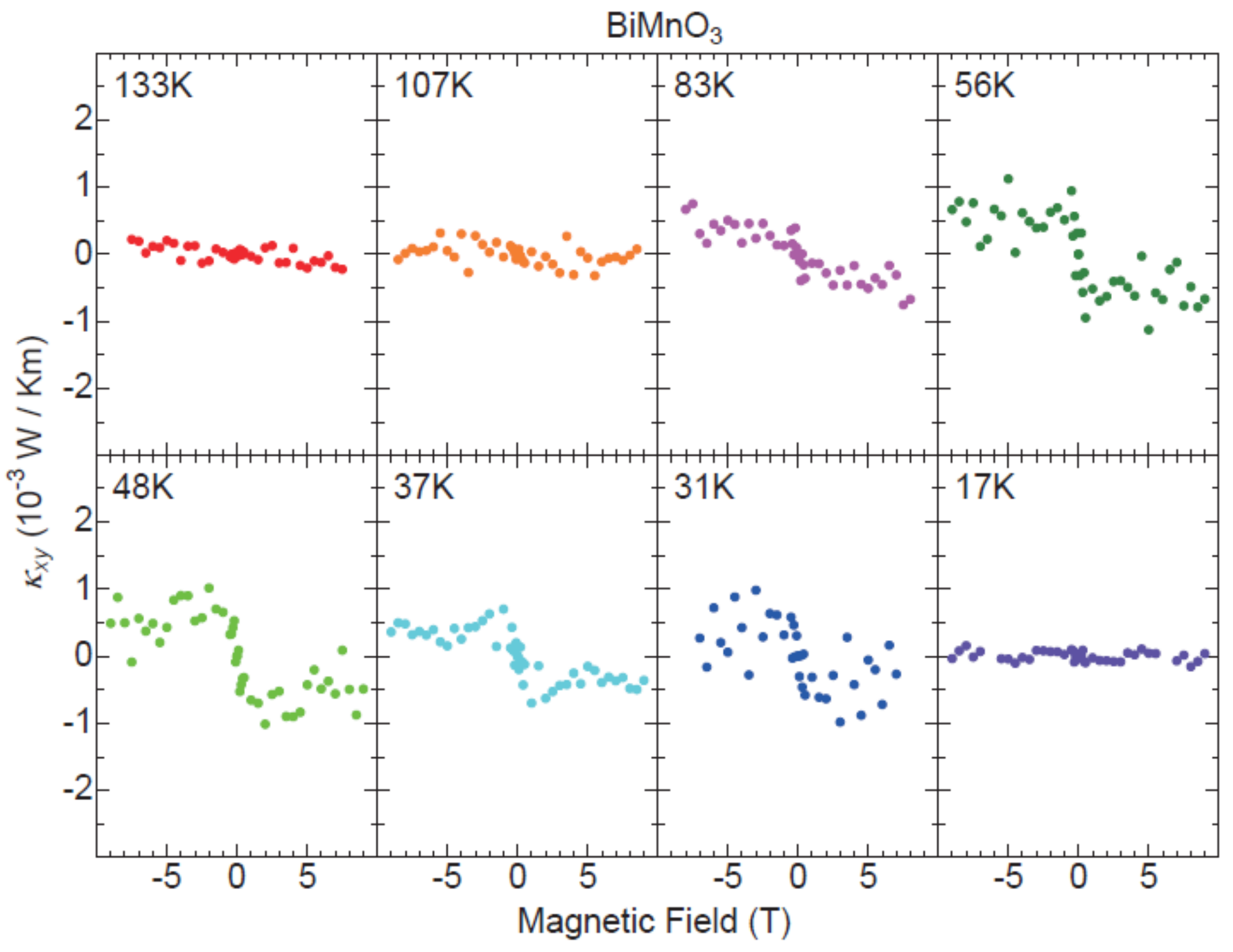}
\caption{(Color online). Magnetic field variation of the thermal Hall conductivity at various temperatures for \BMO.}
\label{fig:12}
\end{center}
\end{figure}

\begin{figure}
\begin{center}
\includegraphics*[width=8.4cm]{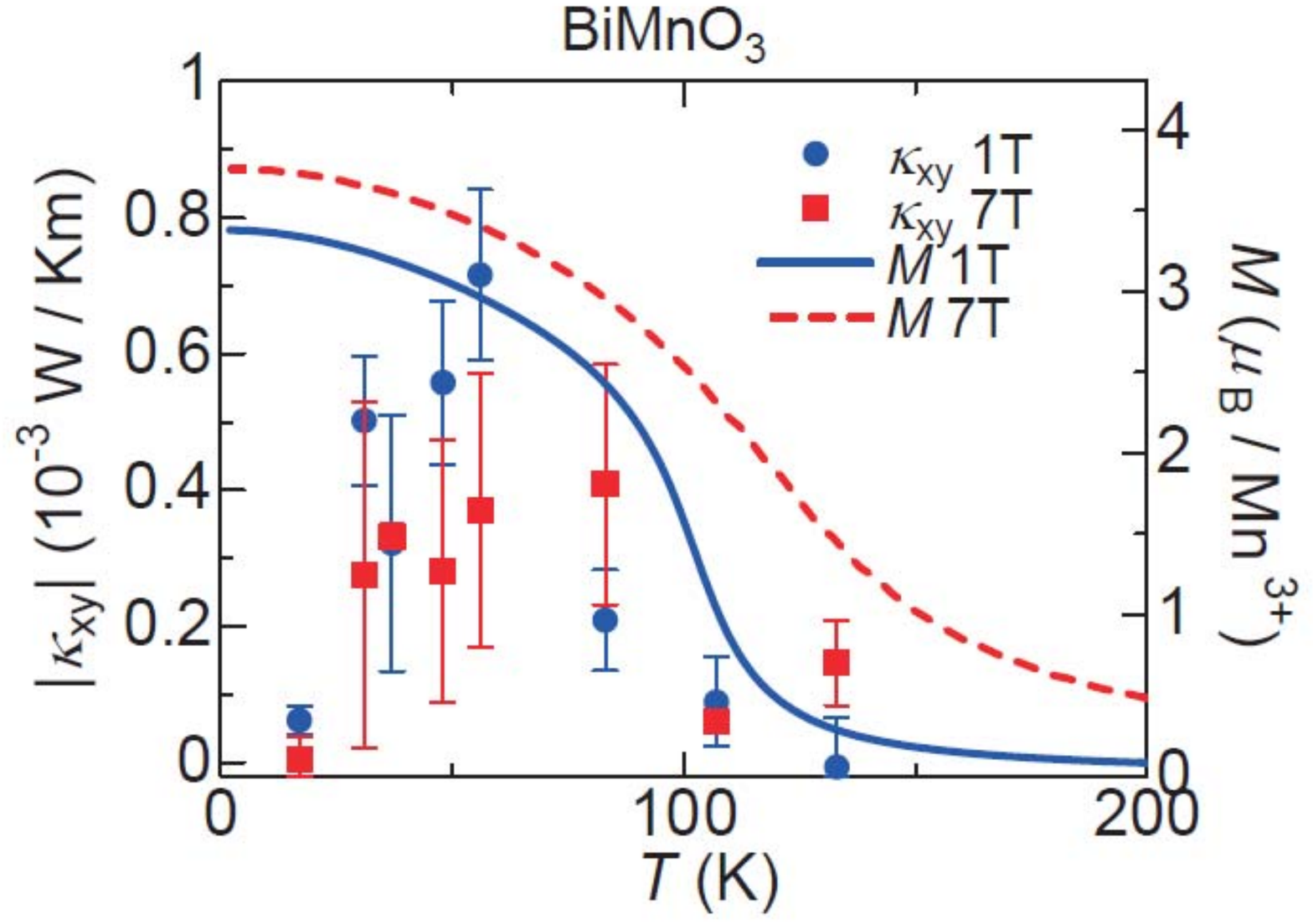}
\caption{(Color online). Temperature dependence of the thermal Hall conductivity and magnetization at $\mu_0 H$ = 1 T and $\mu_0 H$ = 7 T for \BMO.}
\label{fig:13}
\end{center}
\end{figure}

\begin{figure}
\begin{center}
\includegraphics*[width=8.4cm]{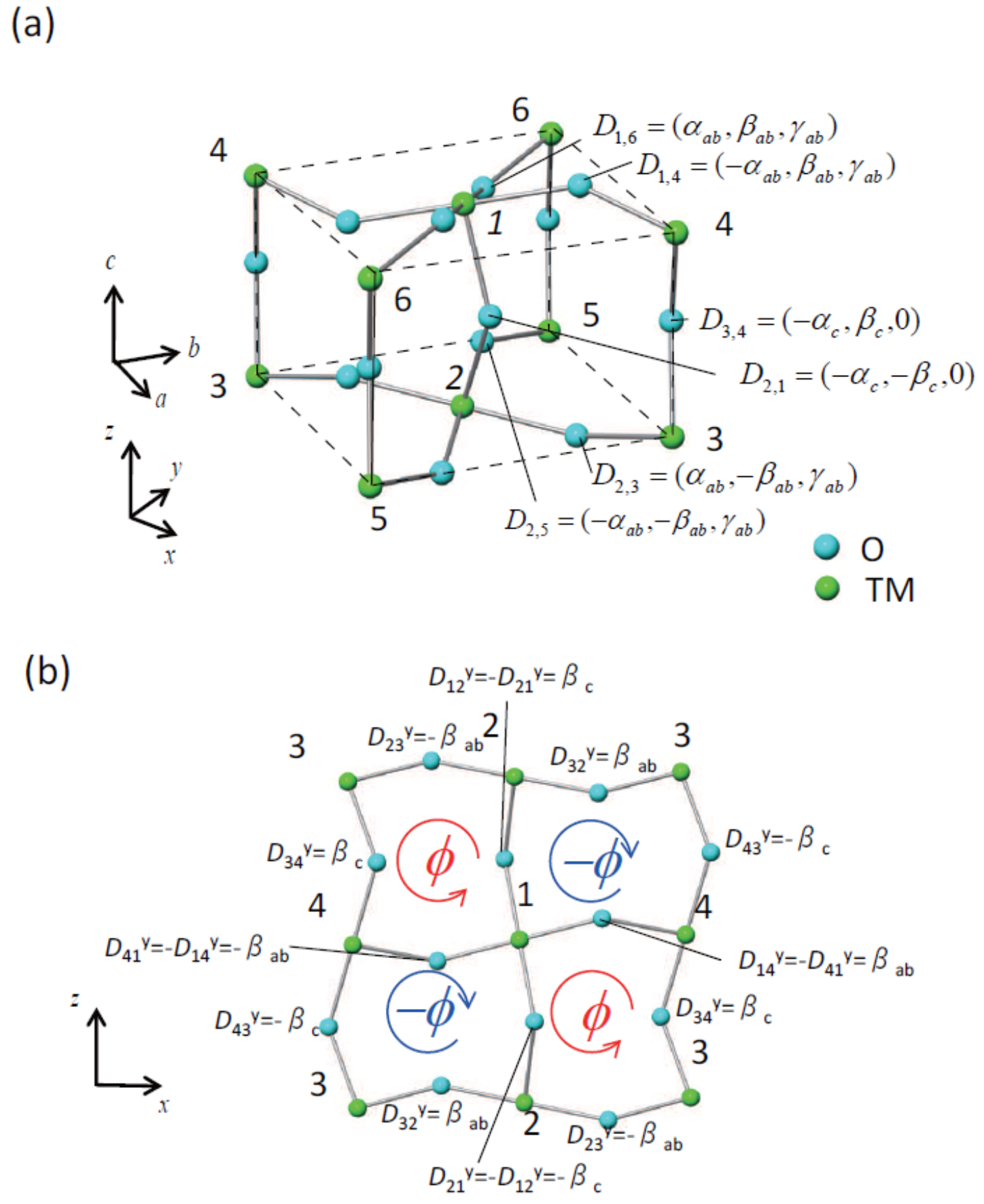}
\caption{(Color online). (a) DM vectors in perovskite structure with GdFeO$_3$-type distortion. For notational convenience, the eqivalent sites 3 and 5 (4 and 6) are distinguished. The $a,b,c$ and $x,y,z$ axes belong to the orthorhombic and pseudocubic coordinate systems, respectively. 
(b) DM vectors and flux pattern in the pseudocubic $zx$ plane. 
The positive direction of flux is taken to be counterclockwise. }
\label{fig:14}
\end{center}
\end{figure}

 In this section, we investigate the magnon Hall effect in ferromagnetic insulators with perovskite like crystal structure
and clarify the effect of lattice geometry on the magnon Hall effect. The crystal structure of perovskite oxide ABO$_3$ with GdFeO$_3$-type orthorhombic distortion is shown in Fig.~\ref{fig:1} (b). 
The crystal structure is composed of corner sharing BO$_6$ octahedra and interstitial A ions. The BO$_6$ octahedra are tilted alternatively in the orthorhombically distorted GdFeO$_3$-type structure. 
The rare earth and transition metal usually occupy the A and B sites, respectively.

The antiferromagnetic interaction usually works between the magnetic moments of nearest neighboring transition metals in perovskite Mott insulators. In order to stabilize the ferromagnetic order in the perovskite lattice, the staggered ordering of two different transition metals is effective. This is certainly realized in double perovskite oxide \LNMO\ , in which Ni$^{2+} (t_{2g}^{6}e_{g}^{2})$ and Mn$^{4+} (t_{2g}^{3})$ ions show the staggered alignment in the B-site sublattice.\cite{La2NiMnO6} The crystal structure is monoclinically distorted with the space group of $P2_{1}/n$ in the low temperature region. According to   
the Kanamori-Goodenough rules, the ferromagnetic superexchange interaction works between Ni and Mn moments. The ferromagnetic spontaneous magnetization is observed below 280 K in La$_2$NiMnO$_6$ as shown in Fig.~\ref{fig:9}(a). The magnetization curve saturates at low magnetic field showing a typical ferromagnetic behavior (Fig.~\ref{fig:9}(b)). 
The saturated magnetic moment is slightly smaller than the expected value (2.5 $\mu_{\rm B}$/(Ni,Mn)).
The resistivity increases rapidly with decreasing temperature (Fig.~\ref{fig:9}(c)).  

Another way to stabilize the ferromagnetic order in perovskite oxides is orbital ordering. 
The orbital-order induced ferromagnetic state is realized in YTiO$_3$ and BiMnO$_3$. YTiO$_3$ has one $3d$ electron per Ti site, which occupies one of the triply degenerate $t_{2g}$ states. The degeneracy is lifted by the four sublattice orbital ordering with d-type Jahn-Teller distortion.\cite{YTiO3-1,YTiO3-2} We show the temperature dependence of magnetization and the magnetization curve at 5 K for YTiO$_3$ in Figs.~\ref{fig:9} (a) and (b), respectively. The ferromagnetic order is stabilized by the orbital ordering below $T_{\rm C}$=30 K. The saturated magnetization almost coincides with the expected value for the $d^1$ electronic configuration (1 $\mu_{\rm B}$/Ti). A neutron diffraction experiment observed the gapless magnon spectrum in this material.\cite{YTiO3-3} The resistivity is too large to be measured even at 300 K, thus not shown in Fig 9(c).

In BiMnO$_3$, four $3d$ electrons at the Mn site shows the high spin configuration due to the large Hund coupling. While three $t_{2g}$ states with the same spin direction are fully occupied, there is only one electron in the doubly degenerated $e_g$ states. As a result, the $e_g$ orbital is ordered with 16 Mn sites in the unit cell as shown in Fig.~\ref{fig:10}.\cite{BiMnO3-1} The temperature dependence of magnetization and the magnetization curve are shown in Figs.~\ref{fig:9} (a) and (b), respectively. The ferromagnetic order is stabilized below around 100 K.\cite{BiMnO3-ferro1,BiMnO3-ferro2} The saturated magnetic moment roughly coincides with the expected value (4 $\mu_{\rm B}$/Mn). The resistivity is too high to be measured also in this case.

Figure~\ref{fig:9} (d) shows the longitudinal thermal conductivity for La$_2$NiMnO$_6$, YTiO$_3$, and BiMnO$_3$. The magnitude of thermal conductivity is small also for these perovskite ferromagnetic insulators. For La$_2$NiMnO$_6$ and YTiO$_3$, the thermal conductivity monotonically decreases with decreasing temperature. On the other hand, it has a broad peak structure around 50 K for \BMO.
 
 We show the thermal Hall conductivity for La$_2$NiMnO$_6$ and YTiO$_3$ in Fig.~\ref{fig:11}, and for BiMnO$_3$ in  Fig. \ref{fig:12}. Finite thermal Hall conductivity is not discernible for \LNMO\ and \YTO\ even below \Tc . On the other hand, we have observed the negative thermal Hall signal below $T_{\rm C}$ in \BMO\ . Similarly to the cases of pyrochlore ferromagnets, the thermal Hall conductivity for BiMnO$_3$ is nearly proportional to the magnetization but tends to decrease with magnetic field after the saturation at low temperature. The temperature dependences of thermal Hall conductivity and magnetization at $\mu_0 H$=1 T and 7 T for BiMnO$_3$ are plotted in Fig.~\ref{fig:13}. 
(Similarly to the previous cases, we plot the averaged thermal Hall conductivity between 0.5 T and 1.5 T as the 1 T value and between 6.5 T and 7.5 T as the 7 T value.) 
The thermal Hall conductivity is suppressed above $T_{\rm C} \approx$ 100 K even at 7 T while the magnetization is large just above $T_{\rm C}$ in the high magnetic field.
Similar to the pyrochlore case, these behaviors can be explained in terms of  magnon Hall effect. 

The presence or absence of the magnon Hall effect in perovskite oxides can also be well explained by the theoretical model based on the DM interaction. In the ideal cubic perovskite structure, where the midpoint between two TM ions is a center of inversion symmetry, the DM interaction is absent. In many materials, however, the DM interaction is allowed because of the distortion, and it is not trivial whether magnon Hall effect can be observed or not. 

As an example, we show the structure and the pattern of DM vectors in \GFO\ type distorted perovskite structure in Fig.~\ref{fig:14}. 
The $a,b,c$ and $x,y,z$ axes are belong to the orthorhombic and pseudocubic coordinate systems, respectively.
In this structure, there are four transition metals (1-4 in the figure) in a unit cell and the DM vectors between the neighboring sites $i$ and $j$ are distributed as follows:
\begin{eqnarray}
\D _{12} &=& (\alpha _{c},\beta _{c},0),\D _{34}=(-\alpha _{c},\beta _{c},0), \\
\D _{23} &=& (\alpha _{ab},-\beta _{ab},\gamma _{ab}),\D _{41}=(\alpha _{ab},-\beta _{ab},-\gamma _{ab}).~~~
\end{eqnarray}
As described in the previous section, only the component of the DM vectors parallel to the magnetic field contributes to the Berry curvature; hence we illustrate the component of DM vectors perpendicular to the lattice plane in Fig.~\ref{fig:14}  (b) ($y$ component in the $zx$ plane).

 In the following discussion, we consider the situation in which the magnetic field is applied along the psudocubic $y$ axis and magnon transfers in the $zx$-plane. When the magnon moves on the loop $1 \rightarrow 2 \rightarrow 3 \rightarrow 4 \rightarrow 1$ in the Fig.\ref{fig:14} (b), magnon gains the phase factor $\phi$ ($\phi$ in the counterclockwise direction), whereas a phase factor is $-\phi$ ($\phi$ in the clockwise direction)
when magnon moves on the next neighboring loop $1 \rightarrow 4 \rightarrow 3 \rightarrow 2 \rightarrow 1$ because $\D _{ij} =-\D _{ji}$. 
Therefore, the flux, which is the phase factor of magnons, are staggered in a $zx$ plane. 
In this case, the absence of the magnon Hall effect can be roughly understood as follows. 
Suppose that the system is purely two-dimensional as described in Fig.~\ref{fig:14}  (b). 
Then if we translate the system by a half length of the lattice constant in the $x$ direction 
and apply the symmetry operation that rotates the plane by angle $\pi$ about the $x$ axis, 
the flux pattern returns to the original one. This immediately implies $\kappa_{zx}=-\kappa_{zx}$ 
and thus the thermal Hall conductivity of magnons is zero. 
Note that the ferromagnetic spin configuration itself is reversed by the above $\pi$ rotation. 
A similar argument applies to the case where the magnetic field is along the $x$ direction or the $z$ direction. 
A more rigorous justification based on the symmetry of 
the (three-dimensional) spin-wave Hamiltonian is given below.

In the following, we explicitly show the cancellation by deriving the effective spin-wave Hamiltonian 
for the case of staggered flux pattern
and showing that the Berry curvature at any ${\bm k}$-point is exactly zero because of symmetry reasons. 
Let us start from the original spin Hamiltonian defined by
\begin{equation}
H = \sum_{\langle i, j \rangle} [J_{ij} {\bm S}_i \cdot {\bm S}_j + {\bm D}_{ij} \cdot ({\bm S}_i \times {\bm S}_j)] -g\mu_{\rm B}H \sum_i S^y_i,
\end{equation}
where $\langle i, j \rangle$ denote nearest neighbor pairs. 
Here, we consider the ferromagnetic exchanges with $J_{ij}=-J_{ab}$ on the $xy$-plane 
and those with $J_{ij}=-J_c$ along the $z$-axis. 
Note that we have neglected further neighbor exchange interactions. 
We then take the $y$-direction as a quantization axis of spins, and apply
the standard Holstein-Primakoff transformation, yielding 
the spin-wave Hamiltonian as
\begin{eqnarray}
H_{\rm SW} &=&
\sum_{\bm k} \Psi^\dagger_{\bm k} {\cal H} ({\bm k})
\Psi_{\bm k},
\\
{\cal H} ({\bm k}) &=& -2S
\begin{pmatrix}
0 & D({\bm k}) \\
D^\dagger (\bm {k}) & 0 \\
\end{pmatrix} + {\rm const.},
\end{eqnarray} 
where $\Psi ({\bm k}) = (b_{1,{\bm k}}, b_{3,\bm k}, b_{2,\bm k}, b_{4, \bm k})^{\rm T}$ and 
\begin{equation}
D ({\bm k}) = \begin{pmatrix}
J_c e^{i\phi} \cos k_z & J_{ab} (\cos k_x + \cos k_y) \\
J_{ab} (\cos k_x + \cos k_y)  &  J_c e^{i \phi} \cos k_z\\
\end{pmatrix}
\end{equation}
with $\phi = \arctan(\beta_c/J_c)-\arctan(\beta_{ab}/J_{ab})$. 
Here we have performed a gauge transformation which makes 
the Hamiltonian less cumbersome.
Note that $k_\alpha = {\bm k} \cdot {\hat \alpha}$ ($\alpha = x,y,z$), 
where ${\hat \alpha}$ corresponds to half the lattice translation in the $\alpha$-direction.

We are now ready to show that the Berry curvature for each band 
at any ${\bm k}$ is exactly zero. 
To this end, we make use of the symmetry of the spin-wave Hamiltonian.  
The Hamiltonian ${\cal H}({\bm k})$ is invariant under the following transformation:
\begin{equation} 
\label{eq:sym_tr}
(\Sigma^x {\cal H}({\bm k}) \Sigma^x)^* = {\cal H} ({\bm k}), 
\end{equation}
where the matrix $\Sigma^x$ is defined by
\begin{equation}
\Sigma^x = 
\begin{pmatrix}
0 & 0 & 1 & 0\\
0 & 0 & 0 & 1\\
1 & 0 & 0 & 0\\
0 & 1 & 0 & 0\\
\end{pmatrix},
\end{equation}
and  asterisk ($*$) denotes the complex conjugate. 
Note that this symmetry is not present in the spin-wave Hamiltonian 
for the pyrochlore ferromagnet Eq. (\ref{eq:sHam_pyro}). 
If we suppose that the eigenvalue of ${\cal H}({\bm k})$ is nondegenerate, 
we can determine the form of the eigenvector that should be invariant 
under the transformation in Eq. (\ref{eq:sym_tr}):
\begin{equation}
\label{eq:eigen_vec}
|\psi_m ({\bm k}) \rangle = [u_m ({\bm k}), v_m ({\bm k}), u^*_m ({\bm k}), v^*_m ({\bm k}) ]^{\rm T},
\end{equation}
where $m$ ($m=1,2,3,4$) are the band indices. 
For the $m$th band, the Berry curvature is given by 
\begin{equation}
\label{eq:Berry_curvature2}
F^{(m)}_{\alpha \beta} ({\rm k}) = -2 {\rm Im} 
\left\langle \frac{\partial \psi_m ({\bm k})}{\partial k_\alpha} \Bigg|
\frac{\partial \psi_m ({\bm k})}{\partial k_\beta} \right\rangle. 
\end{equation}
However, due to the special form of the eigenvector Eq. (\ref{eq:eigen_vec}),  
one can easily see that the right-hand side of Eq. (\ref{eq:Berry_curvature2}) is always zero. 
This proves that $F^{(m)}_{\alpha \beta} ({\bm k}) =0$ for any ${\bm k}$.  
The same symmetry  applies to the case where the magnetic field is 
applied along the $z$-axis or $x$-axis. Thus, we conclude that the thermal Hall conductivity 
$\kappa_{\alpha \beta}$ should vanish in this system irrespective of the direction of the field. 

YTiO$_3$ certainly has the GdFeO$_3$-type orthorhombic crystal structure. While the crystal structure of \LNMO\ is monoclinic, it can be approximately viewed as GdFeO$_3$ structure because the difference of monoclinic angle from 90 degree is less than 0.1 degree.\cite{LNMO2} Therefore, the absence of thermal Hall effect in YTiO$_3$ and La$_2$NiMnO$_6$ is well explained by the above theory. On the other hand, \BMO\ have a larger unit cell with 16 Mn sites. In this case, the cancellation of the Berry curvature due to symmetry reasons may be avoided, and thus a finite thermal Hall conductivity is expected and actually observed.

\section{Summary}

In summary, we have investigated the magnon-induced thermal Hall conductivity in various ferromagnetic insulators. In addition to the previously reported Lu$_2$V$_2$O$_7$ case,\cite{magnon-Hall-exp} the finite thermal Hall conductivity is also observed in other pyrochlore ferromagnets Ho$_2$V$_2$O$_7$ and In$_2$Mn$_2$O$_7$. The temperature and magnetic field dependencies as well as the order of magnitude are quite similar to those of Lu$_2$V$_2$O$_7$, which indicates that the observed thermal Hall conductivity in the pyrochlore ferromagnets can be generically ascribed to the magnon Hall effect due to Dzyaloshinskii-Moriya interaction.
In the perovskite ferromagnets, the thermal Hall signal is indiscernible for  \LNMO\ and \YTO\ , which contains 4 transition metal(TM) sites in a unit cell, but a finite signal is observed for BiMnO$_3$ with the larger unit cell with 16 TM sites. 
The presence or absence of magnon Hall effect can also be well explained by the theory based on the Dzyaloshinskii-Moriya interaction.
Our study further revealed that the lattice geometry affects the topological strucuture of the Bloch wavefunction
and is thus essential for the observed magnon Hall effect.

The authors thank M. Mochizuki, J. Fujioka, N. Kanazawa, and T. Kurumaji for frutiful discussions. This work was supported by a Grant-in-Aid for Scientific Research (nos. 17071007, 17071005, 19048008, 19048015, 19684011, 20046004, 20340086, 21244053, 22014003, 23684023, and 23740298) from the Ministry of Education, Culture, Sports, Science and Technology of Japan and by the Funding Program for World-Leading Innovative R$\&$D on Science and Technology (FIRST), Japan.

\end{document}